\documentclass[aps,prb,reprint,superscriptaddress]{revtex4-1}
\usepackage{amsfonts,amsmath,amssymb,bm}
\usepackage{graphicx}
\usepackage{times,mathrsfs}
\usepackage[colorlinks=true,linkcolor=blue,filecolor=blue,urlcolor=blue]{hyperref}
\usepackage{soul,xcolor}
\usepackage{wasysym}

\begin{document}

\setstcolor{red}

\title{Kinetic frustration induced supersolid in the $S=1/2$ kagome lattice antiferromagnet in a magnetic field}

\author{Xavier Plat}
\affiliation{Condensed Matter Theory Laboratory, RIKEN, Wako, Saitama 351-0198, Japan}
\affiliation{iTHES Research Group, RIKEN, Wako, Saitama 351-0198, Japan}
\author{Tsutomu Momoi}
\affiliation{Condensed Matter Theory Laboratory, RIKEN, Wako, Saitama 351-0198, Japan}
\affiliation{RIKEN Center for Emergent Matter Science (CEMS), Wako, Saitama, 351-0198, Japan}
\author{Chisa Hotta}
\affiliation{Department of Basic Science, University of Tokyo, Tokyo, 153-8902, Japan}

\date{\today}


\begin{abstract}
We examine instabilities of the plateau phases in the spin-1/2 kagome-lattice antiferromagnet in an applied field by means of degenerate perturbation theory, and find some emergent supersolid phases below the $m=5/9$ plateau. The wave functions of the plateau phases in a magnetic field have the particular construction based on the building blocks of resonating hexagons and their surrounding sites. Magnon excitations on each of these blocks suffer from a kinetic frustration effect, namely, they cannot hop easily to the others since the hopping amplitudes through the two paths destructively cancel out with each other. The itineracy is thus weakened, and the system is driven toward the strong coupling regime, which together with the selected paths allowed in real space bears a supersolid phase. This mechanism is contrary to that proposed in lattice-Bose gases, where the strong competing interactions suppress with each other, allowing a small kinetic energy scale to attain the itinerancy. Eventually, we find a supersolid state in which the pattern of resonating hexagons are preserved from the plateau crystal state and only one-third of the originally polarized spins outside the hexagons dominantly join the superfluid component, or equivalently, participate in the magnetization process.
\end{abstract}

\maketitle

%%%%%%%%%%%%%%%%%%%%%%%%%%%%%%%%%%%%%%%%%%%%%%%%%%%%%%%%%%%%%%%%%
%%%%%%%%%%%%%%%%%%%%%%%%%%%%%%%%%%%%%%%%%%%%%%%%%%%%%%%%%%%%%%%%%
\section{Introduction}
\label{sec:intro}
%%%%%%%%%%%%%%%%%%%%%%%%%%%%%%%%%%%%%%%%%%%%%%%%%%%%%%%%%%%%%%%%%
%%%%%%%%%%%%%%%%%%%%%%%%%%%%%%%%%%%%%%%%%%%%%%%%%%%%%%%%%%%%%%%%%

Frustrated systems offer a platform to study the interplay of competing interactions and quantum fluctuations, which has been the source of several concepts including spin liquids,\cite{fazekas1974,anderson1987,Balents2010} valence bond crystals,\cite{read1989} spin nematics,\cite{andreev1984,shannon2006} electronic frustration,\cite{ikeda2000,hotta2012} etc. However, it is difficult to understand precisely how the ``frustration" acts in the respective phases of matter. In overall, geometrical frustration means the competing interactions that leads to a macroscopically quasi-degenerate low-energy structure, which illustrates the impossibility to form trivial types of long-range orders. It is then expected that quantum fluctuations will select particular configurations out of this degenerate manifold, a mechanism called order-by-disorder.~\cite{Villain1980} This mechanism indeed gives a good description of how the supersolids could be formed in Bose gases on optical lattices.\cite{Kovrizhin2005,Goral2002,danshita2009} There, the interactions between boson are large compared to the kinetic energy, and the degenerate manifold of states are mixed but the resultant state still keeps a part of the sites to have a localized component of bosons while the rest of the sites join a superfluid. Supersolid is an established phase of matter,\cite{andreev1969,legget1970,Matsuda1970,Liu1973} but still elusive in a sense that, after the first realistic proposal on solid helium-4\cite{kim2004}, only few examples appeared; only recently that the supersolid proposed in theory\cite{Kovrizhin2005,Goral2002,danshita2009} was found in experiments on atomic quantum gas on optical lattice.\cite{Landig2016} Other than that, there are some in some small windows of the magnetization curves of frustrated quantum spin system with large Ising anisotropy, or equivalently, strongly interacting hard core bosons on the frustrated square\cite{batrouni2000,hebert2001,sengupta2005} and triangular\cite{wessel2005,melko2005,Yamamoto2014} lattices. It also appears in the frustrated Heisenberg antiferromagnets in the triangular and Shastry-Sutherland lattices.~\cite{Momoi2000b}

In quantum magnets, the ``interactions" refer to the Ising type of interactions, and ``quantum fluctuations" or ``kinetic term" to the XY interactions. In the Heisenberg model, the two energy scales are essentially the same, whereas in the XXZ model with strong Ising anisotropy, the strong coupling picture of the aforementioned atomic gas is applied. Thus, even though the supersolids of frustrated quantum magnets look quite alike with those proposed in atomic gasses, the Heisenberg ones are the exceptions that are not due to the interplay of large interactions and frustration. Indeed, a magnetic field already splits these degenerate energy levels into several pieces of manifolds,
and the severe frustration due to competing interaction is partially resolved. In such case, another important aspect of frustration plays a crucial role.

To highlight this point, we here focus on the antiferromagnets on the Shastry-Sutherland and kagome lattices. These two are the limited numbers of nontrivial systems that host numbers of plateau phases; plateaus are the spin gapped phases in an applied field, and many of them can be regarded as crystals of magnons.~\cite{chubokov1991,cabra1997,Totsuka1998,Mila1998,momoi1999,Momoi2000a,Momoi2000b} In this picture, the magnetic field plays the role of a chemical potential of the bosonic magnetic particles. Depending on the relative strength of the kinetic and interaction terms, they generally form either a superfluid (or Bose Einstein condensate (BEC)) or a crystal, \textit{i.e.} a spin density wave phase. The necessary condition to have a plateau is a well established commensurability criterion on the magnon density,~\cite{Oshikawa2000,Hastings2004,Nachtergaele2007} and a large enough interaction gives a sufficiency. 

Now, the above mentioned two lattices afford the strong reduction of the magnon hopping by the {\it kinetic frustration effect},\cite{barford1991,merino2006} called {\it destructive interferences} as well. The Shastry-Sutherland antiferromagnet has an orthogonal dimer lattice with an exact dimer-product ground state.~\cite{Miyahara1999}
In this state, both sites of a dimer are connected to the same nearby site, and that site cannot have a finite hopping weight from the dimer due to the cancellation of
the weight from two paths.~\cite{Miyahara1999} Then, the magnon kinetics is dominated by the hopping processes mediated by another bosons, which is called the correlated hopping.~\cite{Momoi2000a,Momoi2000b} This particular suppression of kinetics against the interaction term favors numbers of plateau phases as the crystals of magnons~\cite{Kageyama1999,Momoi2000a,Momoi2000b,Dorier2008,Sebastian2008} or possibly pairs of bound magnons,~\cite{Corboz2014} and supersolid phases nearby.~\cite{Momoi2000b,takigawa2008} Similar kinetics takes place for the kagome system,\cite{Cabra2005,Capponi2013,Hida2001,Huerga2016,Nakano2010,Nishimoto2013} where the resonating magnons on hexagonal units cannot hop to the surrounding sites due to the cancellation just as those on the dimers of the Shastry-Sutherland model, which we explain in more detail shortly. This kind of kinetic frustration is also well-known as the origin of the flat band in a series of line graphs including the kagome lattice.\cite{mielke1992}

In the present paper, we examine the phases in the magnetization curve of the kagome antiferromagnet off the plateaus, whose nature has not yet been disclosed, and find that the kinetic frustration plays a key role in keeping the magnons relatively localized, which drives the system to a strong coupling region. The phase thus formed in the vicinity of a plateau is a typical {\it supersolid of magnons}.

So far, there are several candidate materials that are considered to have the magnetic systems of kagome geometries, such as ZnCu$_3$(OH)$_6$Cl$_2$,\cite{Bert2007,Asaba2014,Fu2015} BaCu$_3$V$_2$O$_8$(OH)$_2$,\cite{Okamoto2009} and CaCu$_3$(OH)$_6$Cl$_2\cdot$0.6H$_2$O.\cite{Yoshida2017} However, the magnetization curve with successive plateaus found in theory\cite{Nishimoto2013,Capponi2013} has not yet been observed yet, whose exploration is an ongoing issue.

The paper is organized as follows: In Sec.~\ref{sec:perturb}, we present our degenerate perturbation theory on the magnon excitations below the $5/9$ plateau. There, we derive the noninteracting one-magnon effective model, thereby examining the instability taking place by the itinerancy of magnons. We also derive two-magnon interactions.
Section~\ref{sec:ss_struct} is devoted to the analysis of the magnetic structure in the possible supersolid phases, stabilized by the magnon interaction, below the 5/9-plateau phase. We finally briefly mention instabilities in the other plateaus in Sec.~\ref{sec:other_plateaus} and conclude  with some remarks in Sec.~\ref{sec:conclusion}.

%%%%%%%%%%%%%%%%%%%%%%%%%%%%%%%%%%%%%%%%%%%%%%%%%%%%%%%%%%%%%%%%%
%%%%%%%%%%%%%%%%%%%%%%%%%%%%%%%%%%%%%%%%%%%%%%%%%%%%%%%%%%%%%%%%%
\section{Degenerate perturbation theory}
\label{sec:perturb}
%%%%%%%%%%%%%%%%%%%%%%%%%%%%%%%%%%%%%%%%%%%%%%%%%%%%%%%%%%%%%%%%%
%%%%%%%%%%%%%%%%%%%%%%%%%%%%%%%%%%%%%%%%%%%%%%%%%%%%%%%%%%%%%%%%%

We consider the Hamiltonian of the spin-1/2 Heisenberg antiferromagnet in a magnetic field on the kagome lattice,
\begin{equation}
{\cal H}= J \sum_{\langle i,j\rangle} {\bm S}_i\cdot {\bm S}_j - h\sum_j S_j^z,
\label{eq:ham}
\end{equation}
where ${\bm S}_i$ is the spin-1/2 vector operator on site-$i$, $J>0$ the antiferromagnetic exchange, $h$ the external magnetic field, and $\langle i,j\rangle $ denotes the neighboring pairs of sites $i$ and $j$. We define the normalized magnetization per site $m=2\sum_{i} S^{z}_{i}/N$, where $N$ is the number of sites, such that the saturation value is $m=1$.

\begin{figure}
\begin{center}
\includegraphics[width=0.95\columnwidth,clip]{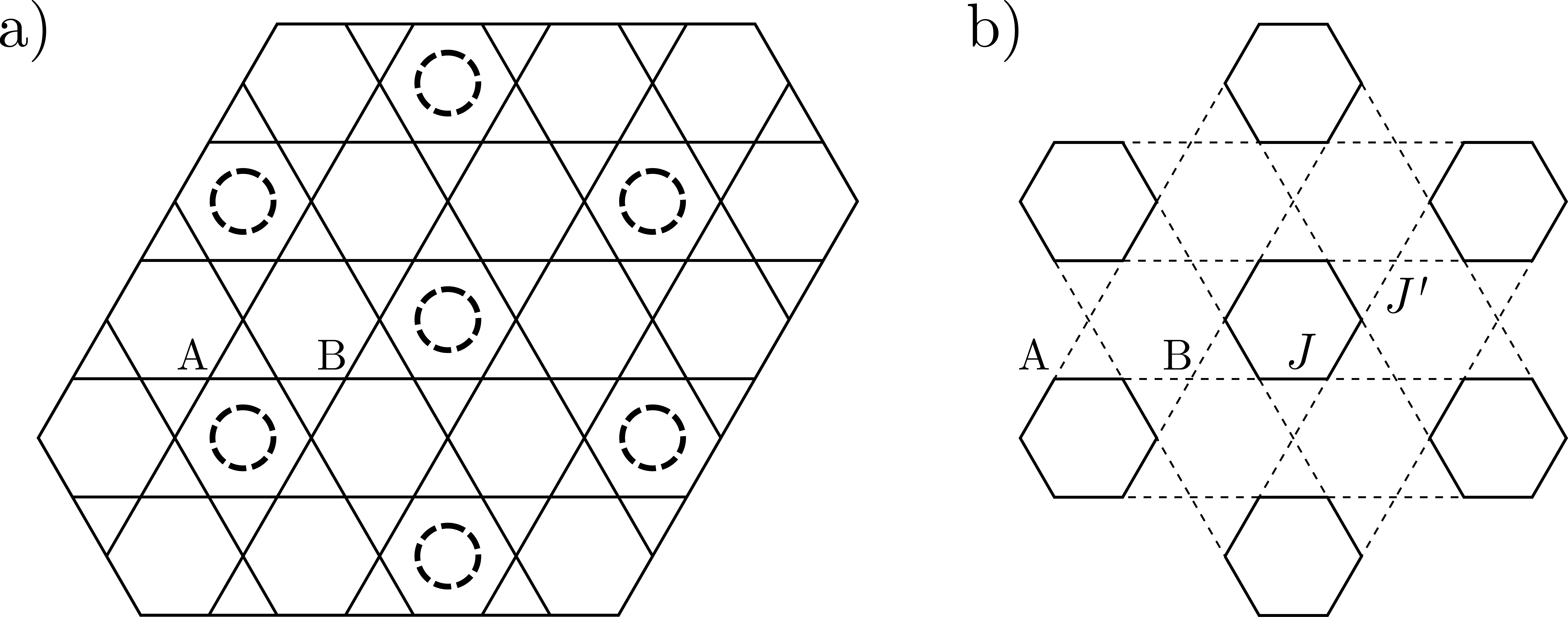}
\end{center}
\caption{(a) $\sqrt{3} \times \sqrt{3}$ structure in the reported  $m=1/3$, 5/9, and 7/9 plateau phases,\cite{Nishimoto2013,Capponi2013}
where hexagons with dashed circles represent localized magnons and the rest of the spins are fully polarized. A and B label those two types of inequivalent sites in the crystal state. (b) Hexamerized kagome lattice used in the perturbation theory approach, which becomes the uniform kagome lattice at $J^\prime=J$. The two lattices have the same space group symmetry.}
\label{fig:lattices}
\end{figure}

Previous studies have revealed that this model has four magnetization plateaus at $m=1/9,1/3,5/9$ and $7/9$, separated by superfluid regions. In the following, we will not discuss the small plateau at $m=1/9$, whose nature is still unclear.~\cite{Nishimoto2013,Picot2016,Huerga2016} There are several numerical evidences that the other plateaus at $m=3/9,5/9,7/9$ host crystals which all have the same symmetry breaking of an extended nine-site unit, called $\sqrt{3}\times\sqrt{3}$ structure.~\cite{Cabra2005,Nishimoto2013,Capponi2013} A schematic picture of those crystals is given in Fig.~\ref{fig:lattices}(a). Hexagons labelled with dashed circles indicate magnons localized on those hexagons, while the rest of the spins, not directly represented here, are fully polarized along the field.

The $m=7/9$ plateau actually provides a particularly clear picture since the product state corresponding to the $\sqrt{3}\times\sqrt{3}$ structure is an exact eigenstate of Eq.~(\ref{eq:ham}).~\cite{Schulenberg2002} Starting from the fully polarized state and introducing a single down spin, one can construct an exactly localized magnon by assigning a staggered phase to the down spin around a hexagon. This results in the cancellation of the hopping outside of the hexagon, which is the aforementioned kinetic frustration effect. We call  the hexagon hosting the localized one-magnon state a \textit{resonating hexagon}. From this single localized magnon, an exact~\cite{Schnack2001,Schmidt2002} ground state for $m=7/9$ is obtained by the tiling the lattice with these resonating hexagons in a fully-packed manner. This state has a finite spin gap and thus leads to the presence of a plateau.~\cite{Momoi2000a,Oshikawa2000} The other two plateaus, although they do not have an exact description, are well approximated by the similar structure of two or three down spins resonating on hexagons and fully polarized spins on all the other sites.~\cite{Nishimoto2013,Capponi2013} Their stability is interpreted in terms of the same mechanism.

%*%*%*%*%*%*%*%*%*%*%*%*%*%*%*%*%*%*%*
%  fig1
%*%*%*%*%*%*%*%*%*%*%*%*%*%*%*%*%*%*%*
%*%*%*%*%*%*%*%*%*%*%*%*%*%*%*%*%*%*%*
%*%*%*%*%*%*%*%*%*%*%*%*%*%*%*%*%*%*%*

In the remaining of this section, we introduce a hexamerized Hamiltonian based on the plateau spatial structure and perform a perturbative expansion of the magnon excitations starting from the plateau state. We focus on the stability of the plateau $m=5/9$ at its low field boundary by examining one-magnon instabilities towards possible supersolid phases.

%%%%%%%%%%%%%%%%%%%%%%%%%%%%%%%%%%%%%%%%%%%%%%%%%%%%%%%%%%%%%%%%%
\subsection{Hexamerized lattice and decoupled limit}
\label{sec:decoupled}
%%%%%%%%%%%%%%%%%%%%%%%%%%%%%%%%%%%%%%%%%%%%%%%%%%%%%%%%%%%%%%%%%

Based on the symmetry breaking pattern of the plateau state, we divide the kagome lattice between the resonating hexagonal units and the surrounding polarized sites, which we denote as A and B, respectively, as shown in Fig.~\ref{fig:lattices}(a). For a lattice of $N$ sites, there are $2N/3$ sites belonging to resonating A-hexagons and $N/3$ B-sites around them.

The starting point of our approach is the aforementioned trial wave function of the $m=5/9$ plateau state, which is a product state given by
\begin{equation}
|\Psi_{5/9} \rangle = \prod_{I} |1,0\rangle_{{\rm A};I} \prod_{i}|\uparrow\rangle_{{\rm B};i},
\label{st:5/9}
\end{equation}
where $|S^z_{\hexagon},l\rangle_{{\rm A};I}$ denotes the $l$-th lowest eigenstate ($l=0,1,2\cdots$) for a given total magnetization $S^z_{\hexagon}$ sector of the Heisenberg model on an \textit{isolated} hexagon, and $|\uparrow \rangle_{{\rm B};i}$ is the $S^z=1/2$ state of the $i$-th B-sites. In the following, we adopt the minuscule and majuscule indices to label the A-hexagons and B-sites, respectively. According to numerical simulations, this wave function gives an accurate description of the plateau state.~\cite{Nishimoto2013,Capponi2013}

\begin{figure}
\begin{center}
\includegraphics[width=0.98\columnwidth,clip]{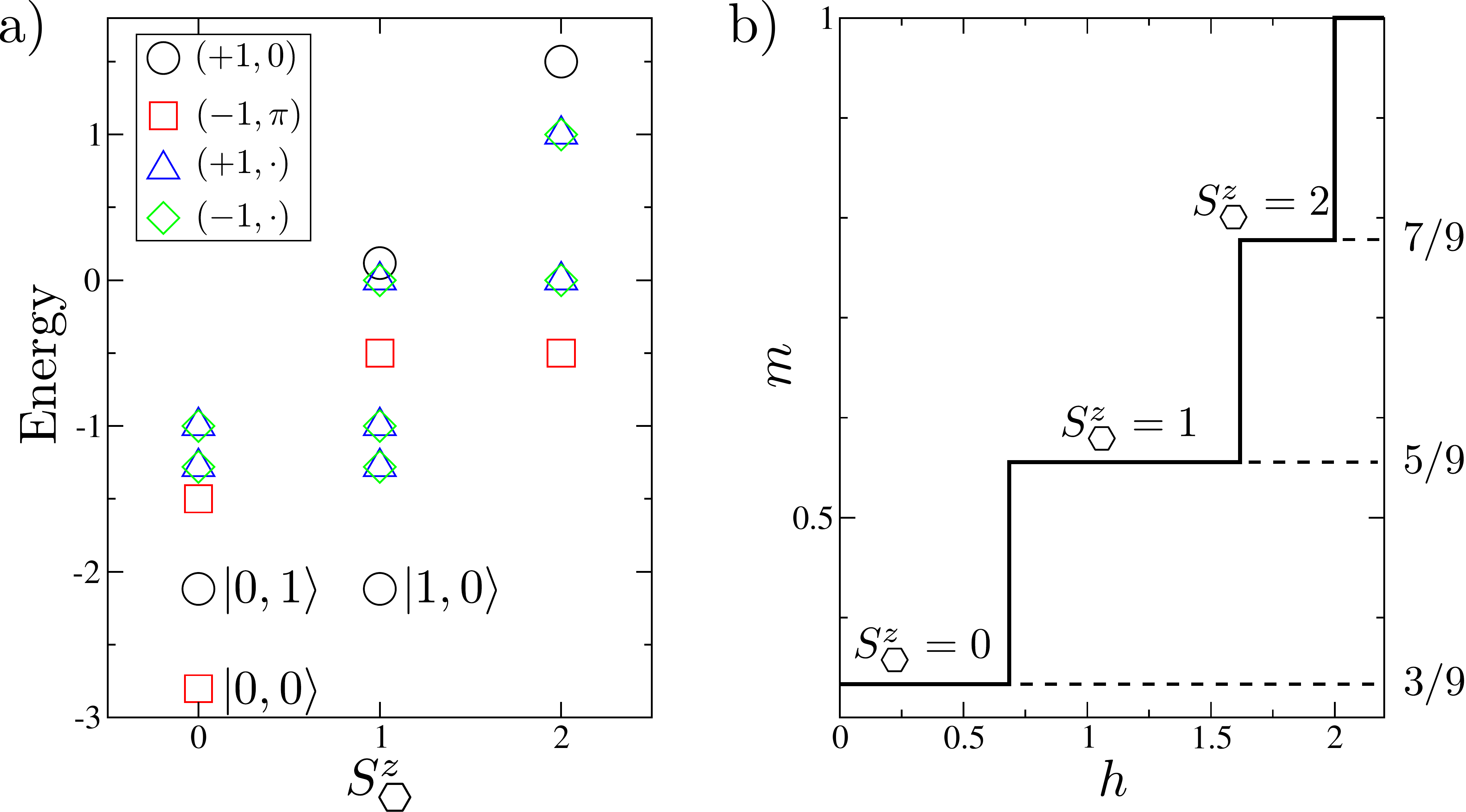}
\end{center}
\caption{(Color online) (a) Lower energy levels of the Heisenberg model on an isolated hexagon in the total magnetization sectors $S^z_{\hexagon}=0,1,2$, with the corresponding eigenstates denoted as $|S^z_{\hexagon}, l\rangle$, with $l=0,\cdots$. The eigenstates are labelled by $(\sigma_{\hexagon},k_{\hexagon})$, where $\sigma_{\hexagon}$ is the reflection symmetry eigenvalue and, when applicable, the momentum eigenvalues $k_{\hexagon}$. (b) Magnetization staircases of the model in Fig.~\ref{fig:lattices}(b) at $J'=0$, with the three plateaux also present in the Heisenberg kagome lattice $J'=J$.}
\label{fig:hex_unit}
\end{figure}

Instead, one can incentively remove the interactions between the A-hexagons and B-spins, 
namely take $J'=0$ in the hexamerized lattice shown in Fig.~\ref{fig:hex_unit}(b). 
In fact, the above trial wave function is the exact ground state for this decoupled hexamerized model.

The excited states from $|\Psi_{5/9} \rangle$ are formed by creating localized magnons on either a A-hexagon or a B-site. 
As we have order-$N$-different choices, each of these excited levels has the degeneracy of that order. 
To prepare those excited states, we need the information on the excitations on the A-hexagons,
namely the eigenstates other than $|1,0\rangle_{{\rm A};I}$.

For this purpose, we report in Fig.~\ref{fig:hex_unit}(a) the lower energy levels obtained 
by the diagonalization of the Heisenberg model on a single hexagon in the sectors $S^{z}_{\hexagon}=0,1,2$. 
The hexagon Hamiltonian is invariant under two symmetry operations, the reflection $\mathcal{R}_{\hexagon}$ about a plane and the translation $\mathcal{T}_{\hexagon}$ along the hexagon. Since those symmetry operations generally do not commute, we can only use one to simultaneously label the eigenstates of the Hamiltonian. In the following, we choose to work with eigenstates of the reflection operation, whose eigenvalues are $\sigma_{\hexagon} = \pm 1$. 
In the case of states with momentum $k_{\hexagon}=0,\pi$, the two operations commute and we also assign them the translation operator eigenvalue. 
The magnetization staircase is shown in Fig.~\ref{fig:hex_unit}(b), 
for which the exact ground states of the form (\ref{st:5/9}) appear in the 1/3, 5/9, and 7/9 plateau states, 
by using the lowest energy states in the sectors $S^{z}_{\hexagon}=0,1,2$, respectively. 

Spins on B-sites are fully polarized as soon as a finite field is introduced.

%%%%%%%%%%%%%%%%%%%%%%%%%%%%%%%%%%%%%%%%%%%%%%%%%%%%%%%%%%%%%%%%%
\subsection{Perturbation overview}
\label{sec:perturb_overview}
%%%%%%%%%%%%%%%%%%%%%%%%%%%%%%%%%%%%%%%%%%%%%%%%%%%%%%%%%%%%%%%%%

The hexamerized decoupled model and the Heisenberg model 
have essentially the same ground state, $|\Psi_{5/9} \rangle$, within a certain range of $h$. 
This motivates us to assign a different interaction $J'$ between A- and B-sites, 
as in Fig.~\ref{fig:lattices}(b), whose associated Hamiltonian $\tilde{\mathcal{H}}$ reads
\begin{equation}
\tilde{\mathcal{H}} = J\sum_{\langle i,j\rangle \in A} {\bm S}_i \cdot {\bm S}_j + J' \sum_{ \langle i \in A, j \in B \rangle} {\bm S}_i \cdot {\bm S}_j -h \sum_j S^z_i,
\label{eq:ham2}
\end{equation}
where the first and second summations run over all nearest neighbor pairs inside the A-hexagons
and on all bonds between A-hexagons and B-sites, respectively.
The kagome lattice Heisenberg model Eq.~(\ref{eq:ham}) is recovered at $J'=J$. 
The coupling $J'$ between the A-hexagons and the B-sites is an appropriate control parameter to understand to how much extent $|\Psi_{5/9} \rangle$
remains stable against varying the magnetic field. 

Following the above context, we adopt a degenerate perturbation theory approach starting from the exact limit $J'=0$.
Throughout this paper, we set $J=1$ as the energy unit and the perturbation strength is measured by the dimensionless coupling $J'$.

As it will be shown shortly, the $|\Psi_{5/9} \rangle$ state has a finite and robust energy gap against 
the fluctuation of the magnon density, which corresponds to the width of the plateau. 
Our objective is to perturbatively calculate the energies of the one-magnon
~\footnote{While the plateau can itself be described as a crystal of magnons, 
from now on the term magnon will refer to states whose magnetization is different from the \unexpanded{$|\Psi_{5/9}\rangle$} state (\textit{e.g.},  one-magnon excitations are obtained by lowering the total magnetization by one starting from \unexpanded{$|\Psi_{5/9}\rangle$})}
excitations above this gap and also their two-magnon interactions.
It is shown that the gap (plateau width) sustains up to $J'=J$, which also indicates the validity of our approach itself.

%*%*%*%*%*%*%*%*%*%*%*%*%*%*%*%*%*%*%*
%*%*%*%*%*%*%*%*%*%*%*%*%*%*%*%*%*%*%*
%  fig2
%*%*%*%*%*%*%*%*%*%*%*%*%*%*%*%*%*%*%*
%*%*%*%*%*%*%*%*%*%*%*%*%*%*%*%*%*%*%*
%*%*%*%*%*%*%*%*%*%*%*%*%*%*%*%*%*%*%*

%*%*%*%*%*%*%*%*%*%*%*%*%*%*%*%*%*%*%*
%  fig3
%*%*%*%*%*%*%*%*%*%*%*%*%*%*%*%*%*%*%*
\begin{figure}
\begin{center}
\includegraphics[width=0.995\columnwidth,clip]{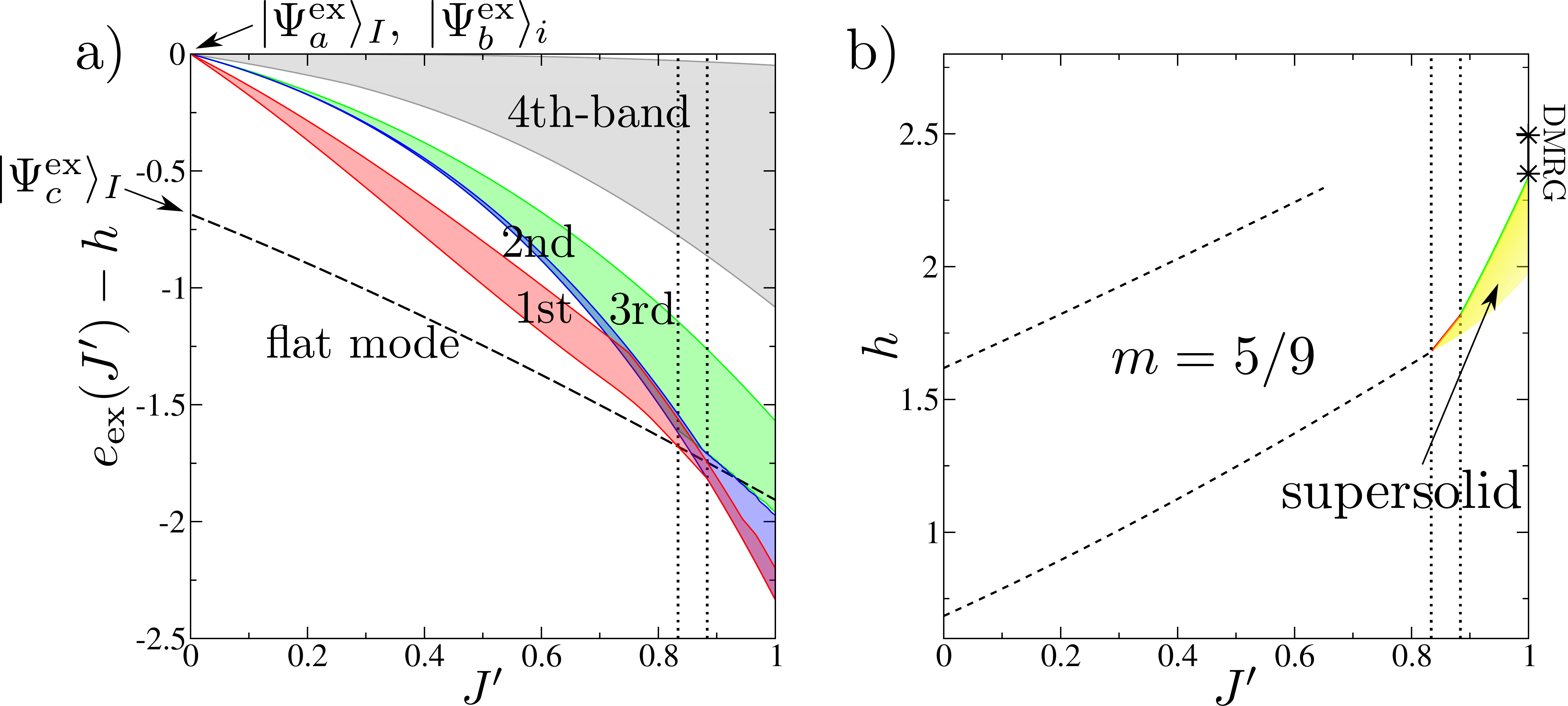}
\end{center}
\caption{(Color online) (a)Evolution of the one-magnon excitation energy, $e_{\rm ex}(J')$, as a function of $J'$, 
measured against the energy of the plateau, $|\Psi_{5/9} \rangle$. 
A magnetic field contributing as a chemical potential of one magnon, $h$, is also subtracted. 
The lower critical phase boundary is hence given by $h_c=e_{\rm ex}(J')$ using 
the energy of the lowest exited state. 
Here, $|\Psi^{\rm ex}_{c} \rangle_I$ and $|\Psi^{\rm ex}_{a/b} \rangle_I$ are
the seeds of the flat mode and the dispersive modes,
respectively. 
The energy range of the four different bands of the dispersive modes are shown. 
(b) Part of the phase diagram we obtained by the perturbation. 
The broken lines are the boundaries of the $m=5/9$ plateau phase evaluated with the flat mode magnon instability. 
The supersolid phase appears when the dispersive mode in (a) replaces the flat mode
and become the lowest one magnon excited state. 
Our scheme cannot identify the lower boundary of the supersolid phase which is left blank, 
and the other plateaus $m=1/3$ and 7/9, are not shown for clarity.
The stars at $J'=1$ mark the region of the 5/9 plateau as obtained by DMRG\cite{Nishimoto2013}. 
}
\label{fig:phased}
\end{figure}
%*%*%*%*%*%*%*%*%*%*%*%*%*%*%*%*%*%*%*

Let us further outline the process of perturbation.
We will consider the one-magnon excitations which decrease the plateau magnetization by one.
Starting from $J'=0$, we track the evolution of these excited levels when $J'$ is turned on, going state-by-state from the lowest ones in energy. In particular, we are interested in how the finite $J'$ will introduce, or not,
a dispersion for the initially localized magnons belonging to the order-$N$ manifold by the mixing of these levels up to the third order perturbation. Figure~\ref{fig:phased} summarizes the development of the two series of one-magnon excited states we studied,
a flat mode and four dispersive modes, as we will develop below. For each $J'$, when starting from the large magnetic field and by decreasing it,
we reach a critical field value $h_c$, where the lowest of the excited energy, $e_{\rm ex}$, becomes zero.
At large enough $J'$, the dispersive modes overwhelm the flat mode and becomes such a lowest excited state 
by gaining kinetic energy as well as a favourable chemical potential. 
This naturally leads to the formation of a supersolid below the plateau, as shown in Fig.\ref{fig:phased}(b).
As we see in the final part of this section, we find that magnons repel with each other, 
which then support the scenario of the continuous transition shown in solid lines in Fig.~\ref{fig:phased}(b).

We note that this approach cannot capture a direct transition to a superfluid phase, which is usually first order
and hence require a drastic change of the wave-function. 
A similar issue was also discussed in the supersolid-solid transition in the XXZ triangular lattice.~\cite{Zhang2011}
In contrast, when the magnons are localized with short-range interactions, 
the transition becomes discontinuous accompanied 
by a macroscopic number of magnon excitations. 
There, a magnetization jump similar to the 
one below the saturation field~\cite{Schulenberg2002,Richter2004,Zhitomirsky2004} 
will be observed.

%%%%%%%%%%%%%%%%%%%%%%%%%%%%%%%%%%%%%%%%%%%%%%%%%%%%%%%%%%%%%%%%%
\subsection{One-magnon instabilities}
\label{sec:magnon_insta}
%%%%%%%%%%%%%%%%%%%%%%%%%%%%%%%%%%%%%%%%%%%%%%%%%%%%%%%%%%%%%%%%%

\subsubsection{Flat band magnon}
\label{sec:flat_band}

At $J'=0$, there are $N_{\mathrm{hex}}=N/9$ degenerate lowest energy one-magnon states obtained by decreasing the magnetization on an $I$-th hexagon. We replace one $|1,0\rangle_{{\rm A};I}$ by $|0,0\rangle_{{\rm A};I}$ as
\begin{equation}
|\Psi^{\rm ex}_{c} \rangle_I = |0,0\rangle_{{\rm A};I} \big(\prod_{J\ne I} |1,0\rangle_{{\rm A};J}\big) \big( \prod_{i} |\uparrow\rangle_{{\rm B};i} \big).
\label{eq:1st_exc}
\end{equation}
By regarding the plateau state as the vacuum, we represent this local excitation by the creation of a hard core boson as $|\Psi^{\rm ex}_{c} \rangle_I =c^{\dagger}_{I}|\Psi_{5/9}\rangle$. To examine its kinetics, we derive the associated effective Hamiltonian up to third order in $J'$ as,
\begin{equation}
\mathcal{H}_{\rm flat} = (-\mu_{\mathrm{loc}}+h) \sum_{I} c^{\dagger}_{I} c_{I},
\label{eq:heff_1mag_loc}
\end{equation}
with $\mu_{\mathrm{loc}} = 0.685 + J^{\prime} + 0.273 J^{\prime 2} - 0.051 J^{\prime 3}$. This effective Hamiltonian measures the excitation energy with respect to the plateau state, \textit{i.e.} it accounts for the energy cost of creating a magnon, which turns out to be localized at this order. This can be straightforwardly understood from the different momenta $k_{\hexagon}=0$ and $\pi$, which results in cancellation of the hoppings through adjacent paths. In fact, because the states $|1,0\rangle_{{\rm A}}$ and $|0,0\rangle_{{\rm A}}$ also have opposite reflection eigenvalues (see Fig.~\ref{fig:hex_unit}(b)), the direct hopping between nearest-neighbour A-hexagons vanishes at any order. Even though longer range hoppings are not constrained to be zero by symmetry properties, they only appear at very high order and we can thus treat this magnon as a localized boson. We here recover the effect of suppression of the kinetic energy scale discussed previously.

As stated above, the $m=1/3$ plateau state at $J'=0$ is a crystal of these localized magnons, $|\Psi_{1/3} \rangle=\prod_I c^\dagger_I |\Psi_{5/9} \rangle$ (we remind it is also a good trial wave function at $J'=1$). However, since they form a flat band, their condensation at the lower critical field of the plateau $m=5/9$ would imply a direct transition from $|\Psi_{5/9} \rangle$ to $|\Psi_{1/3} \rangle$, and thus a magnetization jump between the two plateaux, similar to the magnetization jump between $m=7/9$ and saturation.~\cite{Schulenberg2002,Richter2004,Zhitomirsky2004}
For $J'=0$, this is the magnetization step already displayed in Fig.~\ref{fig:hex_unit}(b), which, setting $J^{\prime}=1$ in (\ref{eq:heff_1mag_loc}), is shifted to $h_{c} \simeq 1.9$.
Since such behaviour has not been reported numerically,~\cite{Nishimoto2013,Picot2016} we need to consider other one-magnon instabilities, which have a higher energy at $J'=0$.

\subsubsection{Dispersive magnons}

Unlike the lowest flat band magnon, the second level of the isolated hexagon in the sector $S^{z}_{\hexagon}=0$, $|0,1\rangle_A$, no longer experiences localization. Indeed, its different quantum numbers, namely $\sigma_{\hexagon}=+1$ and $k_{\hexagon}=0$, do not lead to destructive interferences of the different hopping paths. Because $|0,1\rangle_{{\rm A}}$ and $|1,0\rangle_{{\rm A}}$ are parts of the same triplet (see Fig.~\ref{fig:hex_unit}), this magnon excitation is simply a spin-flip of energy given by $-h$ at $J'=0$. One can alternatively flip one of the polarized spins on a B-site with the same energy cost, and we thus need to consider the following two types of degenerate one-magnon excitations, of degeneracies $N/9$ and $N/3$ respectively,
\begin{equation}
\begin{split}
|\Psi^{\rm ex}_{a} \rangle_I &= |0,1\rangle_{{\rm A};I} \big(\prod_{J\ne I} |1,0\rangle_{{\rm A};J}\big) \big( \prod_{i} |\uparrow\rangle_{{\rm B};i} \big),\\
|\Psi^{\rm ex}_{b} \rangle_i &= \big(\prod_{I} |1,0\rangle_{{\rm A};I}\big) |\downarrow\rangle_{{\rm B};i} \big( \prod_{j \ne i} |\uparrow\rangle_{{\rm B};j} \big).\\
\end{split}
\label{eq:2nd_exc}
\end{equation}

%*%*%*%*%*%*%*%*%*%*%*%*%*%*%*%*%*%*%*
%  fig4
%*%*%*%*%*%*%*%*%*%*%*%*%*%*%*%*%*%*%*
\begin{figure*}
\begin{center}
\includegraphics[width=1.9\columnwidth,clip]{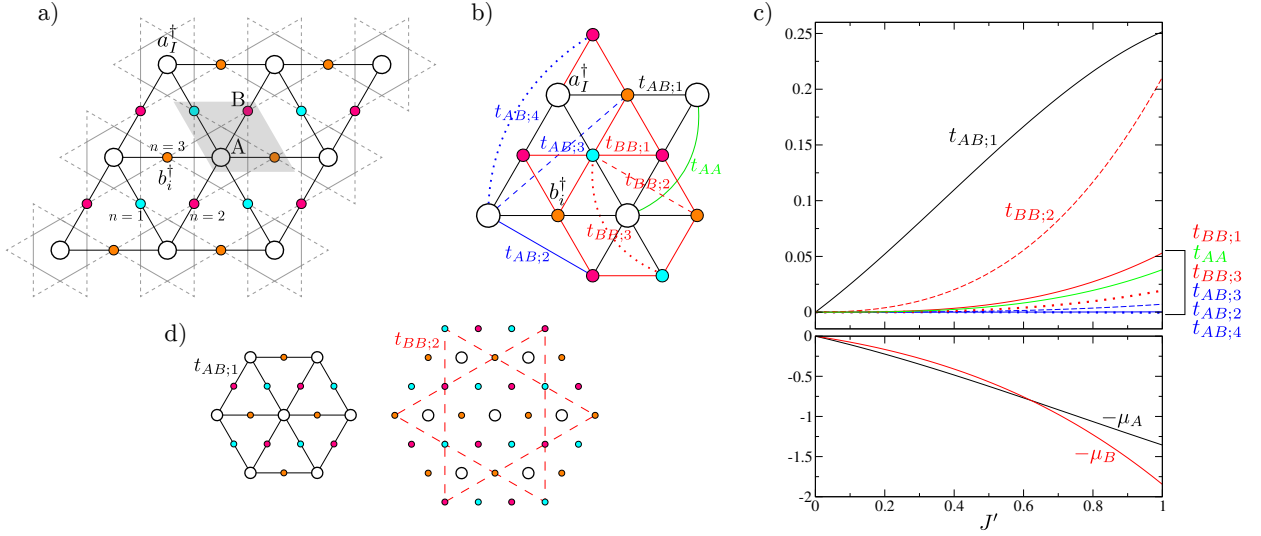}
\end{center}
\caption{(Color online) (a) Effective decorated triangular lattice: the hexagon sites (big empty circles) at the centers of A-hexagons form a triangular lattice hosting the $a^{\dagger}_{I}$ magnons, and the three sublattices of B-sites (filled colored circles) with the $b^{\dagger}_{i}$ magnons give decorating sites in the center of every triangular lattice bond. In grey is the unit cell of the effective lattice. (b) Hoppings connections of the effective model Eq.~(\ref{eq:heff_1mag}). (c) Hoppings amplitudes (top) and chemical potentials (bottom) as functions of $J^{\prime}$. (d) Spatial geometry of the two dominant transfer integrals $t_{AB;1}$ and $t_{BB;2}$. The paths of $t_{BB;2}$ form three decoupled large kagome lattices, of which only one is drawn here.}
\label{fig:heff}
\end{figure*}
%*%*%*%*%*%*%*%*%*%*%*%*%*%*%*%*%*%*%*
%*%*%*%*%*%*%*%*%*%*%*%*%*%*%*%*%*%*%*

It is convenient to introduce two species of bosonic operators, $a^{\dagger}_{I}$ and $b^{\dagger}_{i}$, which create those magnons as $|\Psi^{\rm ex}_{a} \rangle_I=a^{\dagger}_{I}|\Psi_{5/9}\rangle$ and $|\Psi^{\rm ex}_{b} \rangle_i=b^{\dagger}_{i}|\Psi_{5/9}\rangle$ on the $I$-th A-hexagon and $i$-th B-site, respectively. These operators are defined on the decorated triangular lattice shown in Fig.~\ref{fig:heff}(a).
We then derive a one-particle effective model using perturbation theory up to third order in $J'$, and obtain the following Hamiltonian
\begin{equation}
\begin{split}
%\mathcal{H}_{\Delta S^z=-1,2}^{(3)}
\mathcal{H}_{\rm kin}
&= \sum_{r=1}^4
\sum_{\langle I,j \rangle_r} t_{AB;r} ( a_{I} b^{\dagger}_j + \mathrm{h.c.} ) \\
&+  \sum_{r=1}^3 \sum_{\langle i,j \rangle_r} t_{BB;r} ( b^{\dagger}_{i} b_{j} + \mathrm{h.c.} ) \\
&+ \sum_{\langle IJ \rangle} t_{AA} ( a^{\dagger}_{I} a_{J} + \mathrm{h.c.} ) \\
&+ (-\mu_{A}+h) \sum_I a^{\dagger}_{I} a_{I} + (-\mu_{B}+h) \sum_i b^{\dagger}_{i} b_{i},
\end{split}
\label{eq:heff_1mag}
\end{equation}
where $t_{AB;r}$, $t_{BB;r}$, $t_{AA}$ are the effective magnon hoppings among the A-hexagons and B-sites, and $\mu_A$, $\mu_B$ are the effective chemical potentials of the two species of magnons. We label each type of transfer integrals by the index $r$, the distance between the two sites in the pairs of nearest to further neighbours, written for instance as $\langle I,j \rangle_r$. The hoppings paths are drawn in Fig.~\ref{fig:heff}(b). Details on the derivation of this Hamiltonian are presented in Appendix~\ref{app:perturb}, and expression of its parameters in Appendix~\ref{app:params}.

In Fig.~\ref{fig:heff}(c), we plot the hoppings and chemical potentials as functions of $J'$. Two important features are visible. First, there is a very large inequivalency of the chemical potential of the $a^{\dagger}_{I}$ and $b^{\dagger}_{i}$ bosons at large enough $J'$. In our target parameter range of $J'\gtrsim 0.8$, $\mu_{A}$ becomes significantly larger than $\mu_{B}$. By comparing their difference $\simeq 0.4$ with the amplitude of the hoppings, we anticipate that the $b^{\dagger}_{i}$ bosons will eventually be favoured, and $a^{\dagger}_{I}$ bosons will be fully suppressed.
The other important feature is that the geometry of the dominant transfer integrals does not simply follow the geometry given by the largest $t_{AB;1}$ hoppings which form a triangular lattice decorated by the mediating B-sites (left of Fig.~\ref{fig:heff}(d)).
Indeed, the second largest hopping is $t_{BB;2}$, which forms a large kagome lattice made from one-third of the B-sites (right of Fig.~\ref{fig:heff}(d)).
Two other independent kagome lattices are formed in the same manner by the rest of the B-sites.
Those three kagome are then coupled to each other by the smaller $t_{BB;1}$ and $t_{BB;3}$ transfer integrals. For simplicity, the very small hoppings $|t_{AB,2-4}| < 0.007 J^{\prime 3}$ are neglected in the rest of our calculations.

The unit cell of model (\ref{eq:heff_1mag}) consists of four sites, one A-hexagon and three B-sites, which  belong to the three sublattices $\Lambda_n$ [see Fig.~\ref{fig:heff}(a)]. For later convenience, we define the Fourier transforms of the four bosonic operators as
\begin{equation}
\begin{split}
a^\dagger_{\bm{k}} &= \frac{2}{\sqrt{N}}\sum_{I} a_I^\dagger \exp (i {\bm k}\cdot {\bm r}_I),\\
b^\dagger_{n,\bm{k}} &= \frac{2}{\sqrt{N}}\sum_{j\in \Lambda_n} b_j^\dagger \exp (i {\bm k}\cdot {\bm r}_j).
\end{split}
\label{eq:fourier}
\end{equation}

\begin{figure}
\begin{center}
\includegraphics[width=1.0\columnwidth,clip]{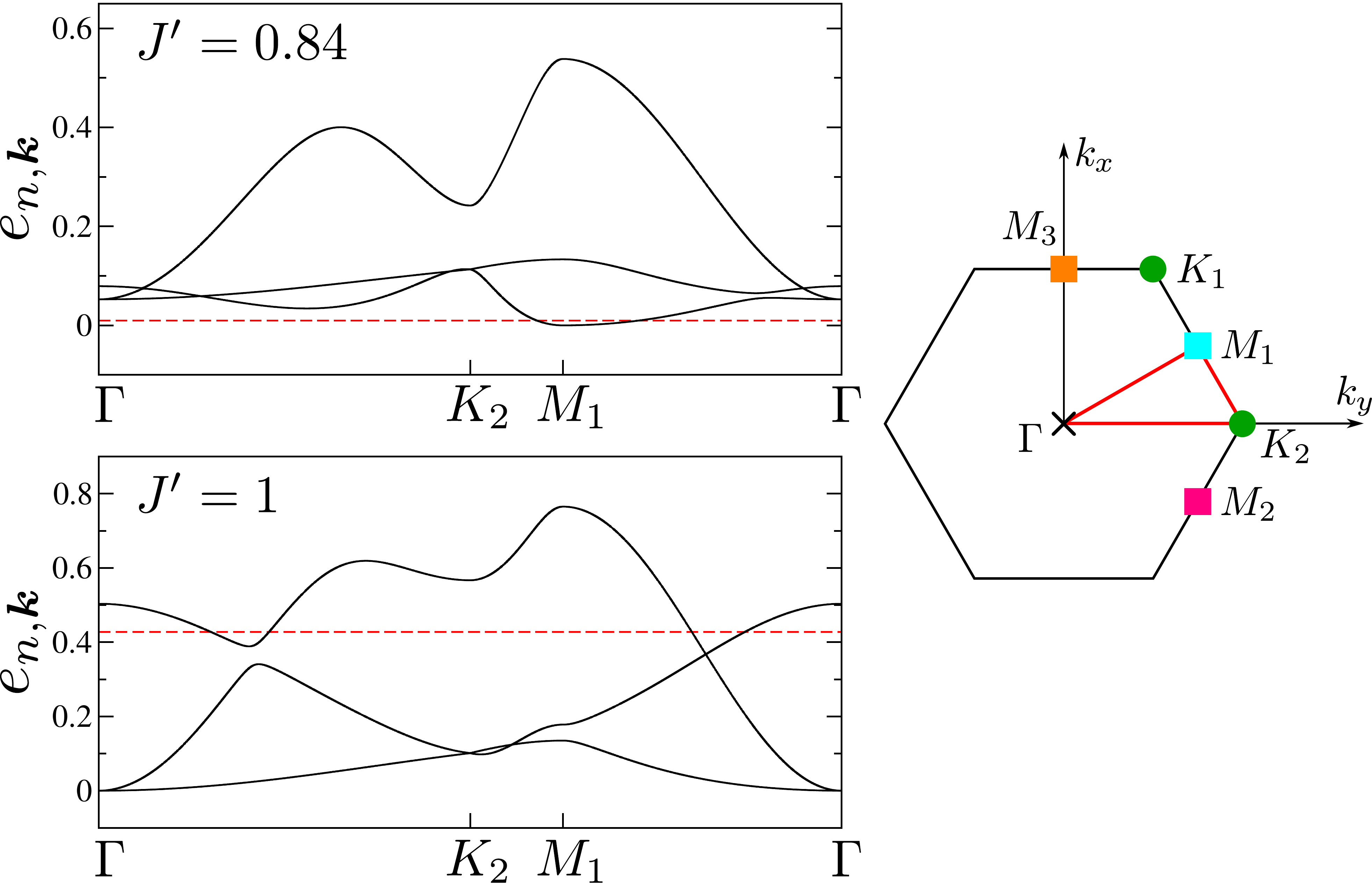}
\end{center}
\caption{(Color online) Left panel: excitation energy spectra of model (\ref{eq:heff_1mag}) for the couplings $J'=0.84$ (top) and $J'=1$ (bottom). The applied magnetic field is chosen so that the lowest energy excitation closes the gap. The solid black lines denote the three lowest bands $e_{n,\bm{k}}$ of the model (\ref{eq:heff_1mag}) along the path $\Gamma K M \Gamma$. The dashed red lines are the flat band coming from the first one-magnon states localized in each hexagon. The minima of these energy bands are the candidates for the one-magnon instability below the $m=5/9$ plateau state, expected to give the lower critical field $h_c$. Right panel: first Brillouin zone of the hexamerized lattice. The coordinates of the high symmetry points are ${\bm K}_{1} = (2\pi/3,2\pi/\sqrt{3})$, ${\bm K}_{2} = (4\pi/3,0)$, ${\bm M}_{1,2} = (\pi,\pm \pi/\sqrt{3})$ and ${\bm M}_{3} = (0,2\pi/\sqrt{3})$. The path used in the left panel is shown in red.}
\label{fig:band_structure}
\end{figure}

We study the band structure of the dispersive model (\ref{eq:heff_1mag}), whose minima give the magnon instabilities. Due to the competition between the hoppings and the chemical potentials, there are four different phases throughout the range $0 < J' \leq 1$, including an incommensurate phase for $0.71 \leq J^{\prime} \leq 0.79$. However, the energy at the minima also has to be compared with the aforementioned flat band, and we find that these minima become the lowest magnon excitation for $J^{\prime} \geq 0.83$.
In this parameter range, we are left with only two phases to investigate, which we call phase I and II.
The three lowest bands are plotted in Fig.~\ref{fig:band_structure}
for two choices of coupling $J^{\prime}=0.84$ and $J^{\prime}=1$ corresponding to phase I and II, respectively,
together with the flat band from the first magnon state.
The one at $J^{\prime} = 0.84$ has minima at the three $\bm M_n$ points at the Brillouin zone boundary, each corresponding to the instability of magnons forming stripes. At $J^{\prime}=1$, the minimum takes place at $\Gamma$-point, where the two bands are degenerate.
For $J'<0.83$, the transition is driven by the localized magnon shown in Sec.~\ref{sec:flat_band},
and hence there is no continuous transition to a supersolid (or simple superfluid) phase
just below the $m=5/9$ plateau phase (see Fig.\ref{fig:phased}), but a jump to the $m=1/3$ plateau.

%*%*%*%*%*%*%*%*%*%*%*%*%*%*%*%*%*%*%*
%  fig5
%*%*%*%*%*%*%*%*%*%*%*%*%*%*%*%*%*%*%*
%*%*%*%*%*%*%*%*%*%*%*%*%*%*%*%*%*%*%*
%*%*%*%*%*%*%*%*%*%*%*%*%*%*%*%*%*%*%*

Before examining those phases in details, let us mention that inclusion of the third order contributions is essential to the above analysis.
At second order, $-\mu_{A}$ is always lower than $-\mu_{B}$, and $t_{AB;1}$ dominates all the other hoppings for any $J'$.
In that sense, the effective model is essentially {\it unfrustrated}, and the model (\ref{eq:heff_1mag}) has only one phase with a unique minimum located at the $\Gamma$ point.
The resulting magnon instability gives a superfluid component with weights $(-\sqrt{3/5},\sqrt{2/15},\sqrt{2/15},\sqrt{2/15})$ on the A-hexagons and the three sublattices of B-sites. Because there is no additional symmetry breaking,
the added component upon the crystal structure is a simple superfluid. The situation drastically changes when third order terms are included, because they introduce a competition among various hoppings at large enough $J'$ (see Fig.~3(b)), and make the chemical potential $-\mu_B$ lower than $-\mu_A$. This change leads to the presence of several different phases which further break lattice symmetries. We note that the frustration effect can be included from the third order perturbation.
Higher-order terms will introduce additional hoppings, which also induce geometrical frustration.

While this can not be taken as a quantitatively accurate value, we also observe that inclusion of the third order terms yields a more precise plateau lower critical field $h_{c} = 2.33$ at $J'=1$, unexpectedly close to the numerics value $\simeq 2.35$.~\cite{Nishimoto2013,Picot2016} 
At second order, the dispersive magnon state achieves a significantly lower energy and the instability thus occurs in a wider range $J' \geq 0.6$, and accordingly, the critical field $h_{c} = 2.7$ underestimates the plateau size.

One can also further continue with the same approach to examine higher one-magnon excitations. For instance, the third lowest magnon state at $J'=0$ is created by changing one hexagon to $|0,2\rangle_{A;I}$. However, its quantum numbers $\sigma_{\hexagon}=-1$ and $k_{\hexagon}=1$ produce localized magnons. Since the effective chemical potential up to third order is small, the associated flat band remains higher in energy than the previous excitations. For even higher magnon states, it is complicated to apply the perturbative scheme because of multiple degeneracies between single hexagon energy levels. Also, they become less clearly well-separated towards the middle of the spectrum, and the validity of the perturbative approach becomes questionable. We have thus limited ourselves to magnon states obtained from the three lowest hexagon states, $|0,l\rangle_{\rm A}$ with $l=0,1,2$.

\subsubsection{Phase I: $M$ points instability}

In the range $0.83 \leq J^{\prime} \leq 0.88$, the lowest band has minima at the three distinct $\bm M_{n}$ points ($n=1,2,3$, see right panel of Fig.~\ref{fig:band_structure}). At these minima, the creation operators of the one-magnon excitations are given by the linear combination of bosons in Eq.~(\ref{eq:fourier})
with wave vector $\bm{k}=\bm{M}_{n}$ as
\begin{equation}
d^{\dagger}_{\bm{M}_{n}} = c_{A} a^{\dagger}_{\bm{M}_{n}} + c_{B,n} b^{\dagger}_{n,\bm{M}_{n}},
\label{eq:creation_Mpoint}
\end{equation}
where the coefficients $c_{A}$ and $c_{B,n}$ continuously depend on $J'$.
The wave vectors $\bm{M}_{n}$ are the three inequivalent center points of the first Brillouin zone boundaries,
shown in Fig.~\ref{fig:band_structure}.
A single $\bm{M}_{n}$ point corresponds to a magnon instability with a doubled unit cell of $2\times4=8$ sites in the direction parallel to $\bm{M}_{n}$.
The distinctive feature of each condensate is that the $d^\dagger_{{\bm M}_{n}}$ magnons have contributions from only one third of the magnons belonging to $\Lambda_n$ in real space, meaning that on the rest of B-sites the spins remain fully polarized. This is interpreted as a supersolid state where the superfluid and the solid components are spatially separated, as illustrated in Fig.~\ref{fig:supersolid_configs}(a) for $\bm{k}={\bm M}_{1}$. At this single-particle level, we cannot anticipate whether the instability will be realized as a condensate at a single $\bm{M}_{n}$ wave vector or as the superposition of more than one condensate. The effect of magnon interactions on this degeneracy will be discussed in Sec.~\ref{sec:magnon_inter}.

We point out that this phase arises from a delicate balance between the effective hoppings (provided that the difference between $\mu_A$ and $\mu_B$ is not too large). Indeed, the two dominant hoppings $t_{AB;1}$ and $t_{BB;2}$ alone cannot explain the minima at the wave vectors $\bm M_n$, and we need to include smaller hoppings. Among them, $t_{AA}$ and $t_{BB:3}$, seem to play an essential role to the formation of the stripes. Since they directly connect the sites belonging to different stripes, even a small amplitude will help this stripe to gain energy against a more complicated phase, e.g. incommensurate $\bm k$-points due to several frustrated hoppings. In terms of $J'$, the variations of the transfer integrals are quite large so that only a small window is allowed for this instability. Thus this phase is somewhat fragile, in the sense that moderate changes to the hoppings at higher orders might mask it.

\subsubsection{Phase II: $\Gamma$ point instability}

For larger coupling $J^{\prime} > 0.88$, the lowest band minimum is located at the $\bm{k}={\bm \Gamma}$ point. As shown on the bottom left panel of Fig.~\ref{fig:band_structure}, there is a band touching at this wave vector. The magnon creation operators at the two-fold degenerate minimum are
\begin{equation}
\begin{split}
d^{\dagger}_{1,\bm{\Gamma}} &= \frac{1}{\sqrt{2}} (b^{\dagger}_{1,\bm{\Gamma}} - b^{\dagger}_{2,\bm{\Gamma}}), \\
d^{\dagger}_{2,\bm{\Gamma}} &=
\frac{1}{\sqrt{6}} (b^{\dagger}_{1,\bm{\Gamma}} + b^{\dagger}_{2,\bm{\Gamma}} - 2 b^{\dagger}_{3,\bm{\Gamma}}).
\end{split}
\label{eq:creation_Gpoint}
\end{equation}
The important feature of these operators is the absence of weight on the $a^{\dagger}_{\bm{\Gamma}}$ bosons. This magnon instability would therefore lead to a supersolid phase where the local magnetizations deviate from the plateau values only on the B-sites, while all the A-hexagons remain in the $|1,0\rangle_{A}$ state and preserve the $\sqrt{3}\times\sqrt{3}$ structure of the plateau. Or, in the bosonic language, the superfluid component only lives on the B-sites.

Contrary to the case of the $\bm{M}_{n}$ instability, this state can be understood from simple considerations.
At large $J^{\prime}$, $-\mu_{B}$ becomes significantly lower than $-\mu_{A}$ by $\simeq 0.5$ (see Fig.~\ref{fig:heff}(c)), leading to the suppression of magnons occupation of the A-hexagons. Consequently, we can keep only the three $t_{BB;r}$ hoppings in the model (\ref{eq:heff_1mag}). The largest one is $t_{BB;2}$ (see Fig.~\ref{fig:heff}(d)), whose geometry produces three independent large kagome lattices, as explained previously. Eventually, the flat bands of the large kagome are coupled by the two smaller couplings $t_{BB;1}$, $t_{BB;2}$ and acquire a finite bandwidth. The band minimum is then simply determined by the largest of those two hoppings, here $t_{BB;1} > 0$. According to this scenario, we believe that the $\bm{\Gamma}$ instability should not be too fragile with respect to higher orders: it mainly relies on the much lower chemical potential level of B-sites, which effectively removes many of the hoppings from the problem, and next on which of $t_{BB;1}$ or $t_{BB;3}$ to be largest other than $t_{BB;2}$.

The physical interpretation of this two-fold degeneracy is that of chiral magnons. This is readily seen by taking the linear combinations
\begin{equation}
\begin{split}
d^{\dagger}_{+,\bm{\Gamma}} & \equiv \frac{i}{\sqrt{2}} d^{\dagger}_{1,\bm{\Gamma}} - \frac{1}{\sqrt{2}} d^{\dagger}_{2,\bm{\Gamma}} \\
&= \frac{1}{\sqrt{3}} ( b^{\dagger}_{3,\bm{\Gamma}} + \omega b^{\dagger}_{1,\bm{\Gamma}} + \omega^2 b^{\dagger}_{2,\bm{\Gamma}} ), \\
d^{\dagger}_{-,\bm{\Gamma}} &\equiv -\frac{i}{\sqrt{2}} d^{\dagger}_{1,\bm{\Gamma}} - \frac{1}{\sqrt{2}} d^{\dagger}_{2,\bm{\Gamma}} \\
&= \frac{1}{\sqrt{3}} ( b^{\dagger}_{3,\bm{\Gamma}} + \omega^2 b^{\dagger}_{1,\bm{\Gamma}} + \omega b^{\dagger}_{2,\bm{\Gamma}} ),
\end{split}
\label{eq:creation_Gpoint_chir}
\end{equation}
where $\omega = \mathrm{exp}(i 2\pi /3)$. These new operators satisfy the chiral relation ${\mathcal{R}}d^{\dagger}_{+,\bm{\Gamma}}=d^{\dagger}_{-,\bm{\Gamma}}$ under the lattice reflection operation $\mathcal{R}$, where the mirror plane is located through bond centers of hexagons.

We notice a close analogy with the magnon instability at the saturation field in the triangular antiferromagnet (TAF).~\cite{Nikuni1995} In this problem, there are also two different lowest energy one-magnon excitations from the fully polarized state
at different wave vectors $\bm{Q}$ and $-\bm{Q}$ on the corners of the triangular-lattice Brillouin zone (the equivalent of the $K_1$ and $K_2$ points in Fig.~\ref{fig:band_structure}). The magnon operators can be expressed as separate contributions from three sublattices of the triangular lattice, $\tilde{\Lambda}_{n}$, as
\begin{equation}
\begin{split}
S^{-}_{\bm{Q}} &= \frac{1}{\sqrt{N}} \left( \sum_{j \in \tilde{\Lambda}_{1}} S^{-}_{j} + \sum_{j \in \tilde{\Lambda}_{2}} \omega S^{-}_j + \sum_{j \in \tilde{\Lambda}_{3}} \omega^2 S^{-}_j \right), \\
S^{-}_{-\bm{Q}} &= \frac{1}{\sqrt{N}} \left( \sum_{j \in \tilde{\Lambda}_{1}} S^{-}_{j} + \sum_{j \in \tilde{\Lambda}_{2}} \omega^2 S^{-}_j + \sum_{j \in \tilde{\Lambda}_{3}} \omega S^{-}_j \right),
\end{split}
\label{eq:creation_triangular_2}
\end{equation}
in which we recognize the same form as Eq.~(\ref{eq:creation_Gpoint_chir}).

%*%*%*%*%*%*%*%*%*%*%*%*%*%*%*%*%*%*%*
%  fig6
%*%*%*%*%*%*%*%*%*%*%*%*%*%*%*%*%*%*%*
\begin{figure}
\begin{center}
\includegraphics[width=\columnwidth,clip]{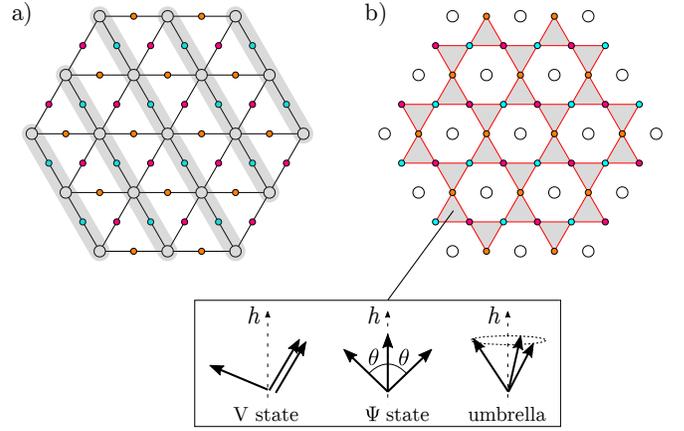}
\end{center}
\caption{(Color online) Superfluids induced by the one-magnon instabilities below the $m=5/9$ plateau at (a)the $\bm{M}_{1}$ and (b) $\bm{\Gamma}$ wave vectors. The grey regions correspond to the sites with finite densities of one magnon species. Inset: the three degenerate configurations of the B-site spins with a three-sublattice structure at the kagome lattice $J^{\prime}=1$ ($\bm{\Gamma}$ instability).}
\label{fig:supersolid_configs}
\end{figure}
%*%*%*%*%*%*%*%*%*%*%*%*%*%*%*%*%*%*%*
%*%*%*%*%*%*%*%*%*%*%*%*%*%*%*%*%*%*%*

Further using the information from the TAF studies, we anticipate that these magnon creation operators can represent three different spin structures:~\cite{Nikuni1995,Yamamoto2014} the so-called $\mathrm{V}$ (or 0-coplanar), $\Psi$ (or $\pi$-coplanar) and umbrella states, shown in Fig.~\ref{fig:supersolid_configs}(b).
In the case of the TAF realized in the XXZ model, magnon interactions select one of the three states due to the spin anisotropy. In a mean-field (MF) approximation, an easy axis (plane) anisotropy selects the $\mathrm{V}$ (umbrella) state.~\cite{Miyashita1986,Watarai2001} Quantum fluctuations somewhat modify the phase diagram by inducing the appearance of a new $\Psi$ phase~\cite{Starykh2014,Yamamoto2014,Sellmann2015,Yamamoto2017} In the next Sec.~\ref{sec:magnon_inter}, we derive magnon-magnon interactions generated by perturbative processes, and in Sec.~\ref{sec:ss_struct} we proceed to a MF calculation to investigate which spin configuration is selected by magnon interactions in our case.

%%%%%%%%%%%%%%%%%%%%%%%%%%%%%%%%%%%%%%%%%%%%%%%%%%%%%%%%%%%%%%%%%
\subsection{Magnon interactions}
\label{sec:magnon_inter}
%%%%%%%%%%%%%%%%%%%%%%%%%%%%%%%%%%%%%%%%%%%%%%%%%%%%%%%%%%%%%%%%%

Since the two types of one-magnon instabilities studied in Sec.~\ref{sec:magnon_insta} had degeneracies, we need to include an extra energy scale to resolve them. The interactions between (dilute) magnons will take care of this role. In the following, we compute these interactions by considering excited states with two magnons and by applying degenerate perturbation theory.

\begin{figure*}
\begin{center}
\includegraphics[width=1.9\columnwidth,clip]{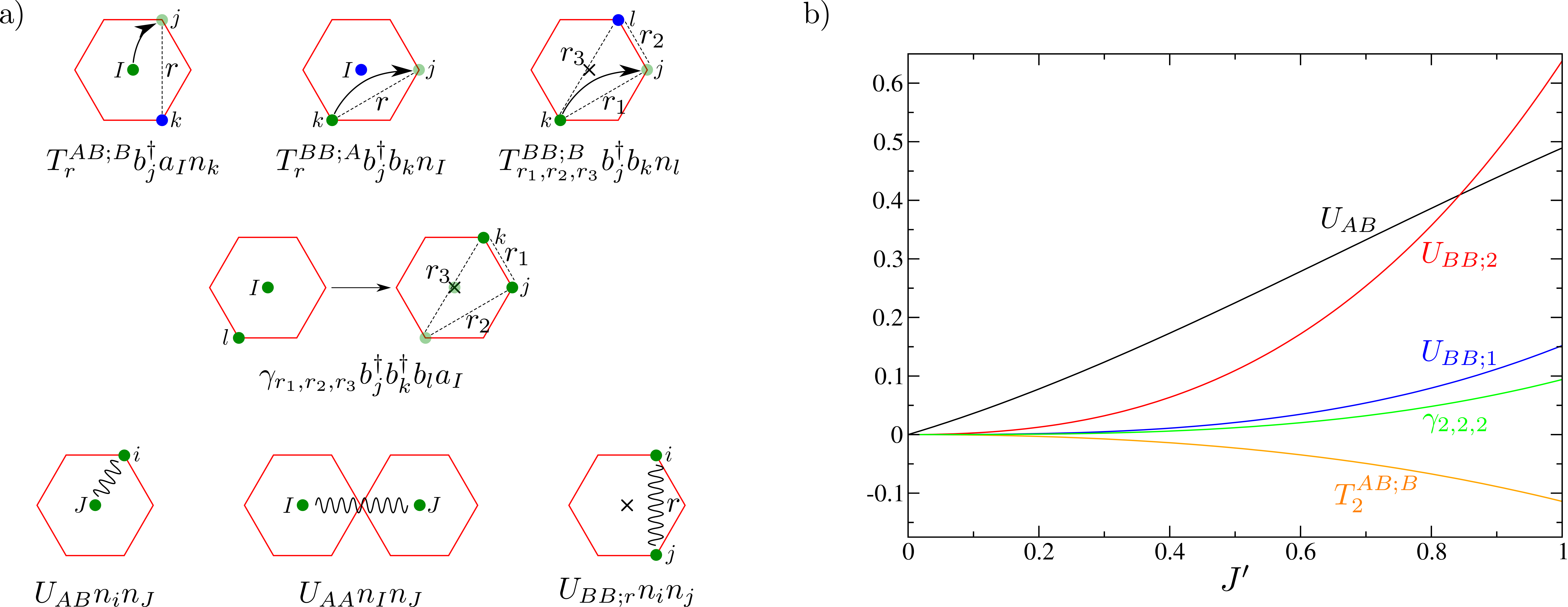}
\end{center}
\caption{(Color online) (a) Illustration of all the two-magnon processes appearing in the effective Hamiltonian (\ref{eq:heff_2mag_proc}). The top row represents the assisted hoppings, the middle row the pair hopping, and the bottom row the density-density interaction terms. Each red hexagon denotes the closed loop of the six B-sites surrounding one A-hexagon (see Fig.~\ref{fig:heff}(b)). Green and blue dots represent the magnons (the latter is used for the assisting magnons of top row). (b) $J'$ dependence of the amplitudes of the five most dominant processes.}
\label{fig:twomag_procs}
\end{figure*}

We restrict our analysis to the evaluation of two-body interactions by working on the magnetization sector with two magnons, $\Delta S^z = -2$. As discussed shortly below, we do not include the states where two magnons exist within the same hexagon. Up to third order, we obtain the interaction terms
\begin{equation}
\mathcal{V}_{\rm int} =V_{\mathrm{assist}} + V_{\mathrm{pair}} + V_{\mathrm{repl}},
\label{eq:heff_2mag}
\end{equation}
where
\begin{equation}
\begin{split}
V_{\mathrm{assist}} &=
\sum_{I} \left[  \sum_{r=1}^{3}\sum_{\langle jk \rangle_r \in {\hexagon_I}} \big\{ T^{AB;B}_r ( b^{\dagger}_{j} a_{I} n_{k} + \mathrm{h.c.} ) \right. \\
&+ %\sum_{r=1}^{3}\sum_{\langle jk \rangle_r \in {\hexagon_I}}
T^{BB;A}_r ( b^{\dagger}_{j} b_{k} n_{I} + \mathrm{h.c.} ) \big\} \\
&+ \left. \sum_{jkl\in\hexagon_I} T^{BB;B}_{r_1,r_2,r_3} ( b^{\dagger}_{j} b_{k} n_{l} + \mathrm{h.c.} ) \right], \\
V_{\mathrm{pair}} &= \sum_{I} \sum_{jkl\in\hexagon_{I}}
 \gamma_{r_1,r_2,r_3} ( b^{\dagger}_{j} b^{\dagger}_{k} b_{l} a_{I} + \mathrm{h.c.} ), \\
V_{\mathrm{repl}} &= U_{AB} \sum_{\langle iJ \rangle} n_{i} n_{J} + \sum_{r=1}^{3} U_{BB;r} \sum_{\langle ij \rangle_r} n_{i} n_{j} \\
&+ U_{AA} \sum_{\langle IJ \rangle} n_{I} n_{J}.
\end{split}
\label{eq:heff_2mag_proc}
\end{equation}
Those three terms respectively contain assisted hoppings, pair hoppings, and repulsive density interactions. The details of the processes, including the definitions of the indices $r_1,r_2,r_3$, are given in Fig.~\ref{fig:twomag_procs}(a). Explicit forms of the non-negligible coefficients $T$, $\gamma$, and $U$ are listed in Appendix~\ref{app:params}, and the amplitudes of the five largest processes as functions of $J'$ are plotted in Fig.~\ref{fig:twomag_procs}(b). In the regime of low magnon density, the probability of having more than three magnons nearby is small; the validity of this effective model thus extends to lower magnetization sectors.

%*%*%*%*%*%*%*%*%*%*%*%*%*%*%*%*%*%*%*
%  fig7
%*%*%*%*%*%*%*%*%*%*%*%*%*%*%*%*%*%*%*
%*%*%*%*%*%*%*%*%*%*%*%*%*%*%*%*%*%*%*
%*%*%*%*%*%*%*%*%*%*%*%*%*%*%*%*%*%*%*

Several simplifications have been made to derive the interaction terms. First, we discarded processes of longer ranges which appear only at third order and of small amplitudes $\lesssim 0.01 J'^3$. Second, we did not take into account the localized magnons (flat band) $c^{\dagger}_{I}$, which should however be included at high magnon density in order to recover the $m=1/3$ plateau. Third, we also ignored all the processes involving an additional type of bosons. In the sector $\Delta S^z = -2$, there is another degenerate local excitation where a single hexagons is excited to $|-1,0\rangle_A$. This local excitation, interpreted as a bound pair of magnons on an A-hexagon, is however dispersionless up to third order and can only decay into two neighbouring magnons $a_I^\dagger$ and $b_i^\dagger$. Because of the large repulsive interaction $U_{AB}$, this situation is unfavorable and the process is safely neglected at low densities.

Finally, we explain that the model (\ref{eq:heff_2mag}) precludes the formation of two-magnon bound states or phase separation. Indeed, all the density-density interactions in (\ref{eq:heff_2mag_proc}) are repulsive and prevent the $a^{\dagger}_{I}$ and $b^{\dagger}_{i}$ magnons to attract with each other. We also verified that inclusion of the flat band does not allow a mechanism in which two neighbouring localized magnons (or a localized magnon $c^{\dagger}_{I}$ and a $b^{\dagger}_{i}$ magnon in our case) can form an itinerating bound pair by means of assisted hopping processes, like in the Shastry-Sutherland lattice.~\cite{Momoi2000b}

%%%%%%%%%%%%%%%%%%%%%%%%%%%%%%%%%%%%%%%%%%%%%%%%%%%%%%%%%%%%%%%%%
%%%%%%%%%%%%%%%%%%%%%%%%%%%%%%%%%%%%%%%%%%%%%%%%%%%%%%%%%%%%%%%%%
\section{Supersolid structures below $m=5/9$}
\label{sec:ss_struct}
%%%%%%%%%%%%%%%%%%%%%%%%%%%%%%%%%%%%%%%%%%%%%%%%%%%%%%%%%%%%%%%%%
%%%%%%%%%%%%%%%%%%%%%%%%%%%%%%%%%%%%%%%%%%%%%%%%%%%%%%%%%%%%%%%%%

In this section, we take into account magnon interactions and show how supersolid phases develop below the $m=5/9$ plateau from the magnon instabilities presented in Sec.~\ref{sec:magnon_insta}.

%%%%%%%%%%%%%%%%%%%%%%%%%%%%%%%%%%%%%%%%%%%%%%%%%%%%%%%%%%%%%%%%%
\subsection{Effective spin model}
%%%%%%%%%%%%%%%%%%%%%%%%%%%%%%%%%%%%%%%%%%%%%%%%%%%%%%%%%%%%%%%%%

To elucidate the spin structures described with the effective Hamiltonian
\begin{equation}
{\cal H}_{\rm eff}={\cal H}_{\rm kin}+{\cal V}_{\rm int},
\label{eq:H_eff_int}
\end{equation}
given in Eqs.~(\ref{eq:heff_1mag}) and (\ref{eq:heff_2mag}), we start by rewriting the model in terms of an effective spin Hamiltonian. We use spin-1/2 pseudo-spin operators to express the degrees of freedom on the A-hexagons as $T^z_I:=1/2 -a^{\dagger}_{I} a_{I}$ and $T^{-}_{I}:=a^{\dagger}_{I}$, whereas the bosons on the B-sites are translated back to the original spin operators $S^z_i=1/2 -b^{\dagger}_{i} b_{i}$ and $S^{-}_{i}=b^{\dagger}_{i}$. Using these spin operators, the model (\ref{eq:H_eff_int}) is expressed in the spin language. We refer to Appendix~\ref{app:mf_ham} for the expression of the spin Hamiltonian and its couplings.

Among various competing interactions, the two dominant couplings are the nearest neighbour spin exchange $J_{AB}$ between A-hexagons and B-site spins and the second neighbor spin exchange $J_{BB;2}$ between B-site spins. Here, we follow the same rules of the indices as those used for the transfer integrals in Eq.~(\ref{eq:heff_1mag}) (see also Fig.~\ref{fig:heff}(b)), while $J_{AB}$ solely denotes the nearest neighbor ones. As previously, the couplings $J_{AB}$ form a decorated triangular lattice and the couplings $J_{BB;2}$ produce three layers of large decoupled kagome lattices given in Fig.~\ref{fig:heff}(d). Both of them are antiferromagnetic, with strong Ising anisotropy. There is also non-negligible antiferromagnetic coupling $J_{BB;1}$ between the nearest neighbor pairs of B-site spins, with a weak XY anisotropy. We notice that the direct exchange interactions between the B-site spins, absent in the original model (\ref{eq:ham2}), is induced by the presence of the resonating hexagons (A-sites), which stabilizes the quantum mechanical spin ordering, as shown in the following section.

The local spin expectation values ${\bm m}_{p}=(m^x_{p},m^y_{p},m^z_{p})$ of the $p$-site belonging to the $I$-th A-hexagon ($p\in\hexagon_I$)
are given by
\begin{align}
  m^\alpha_p & = \frac{\sqrt{2}}{6} \langle T^\alpha_I \rangle \ \ \ \ (\alpha=x,y),\nonumber\\
  m^z_p & = \frac{1}{6}\left( \langle T^z_I \rangle + \frac{1}{2} \right),
  \label{eq:mag_h}
\end{align}
and the spin expectation values on the three B-site sublattices are directly given by the original $\langle {\bm S}_i \rangle$,
\begin{align}
  m^\alpha_i & = \langle S^\alpha_i \rangle,
  \label{eq:mag_ss}
\end{align}
($\alpha=x,y,z$) for $i\in\Lambda_n$.

%%%%%%%%%%%%%%%%%%%%%%%%%%%%%%%%%%%%%%%%%%%%%%%%%%%%%%%%%%%%%%%%%
\subsection{Mean field approximation}
%%%%%%%%%%%%%%%%%%%%%%%%%%%%%%%%%%%%%%%%%%%%%%%%%%%%%%%%%%%%%%%%%

We now apply a MF approximation to the effective spin model (\ref{eq:heff_2mag_spin}). Then, we take the classical limit, replacing all spin-1/2 vector operators with classical vectors of length 1/2. Assuming a unit cell consisting of four sites involving one A-hexagon and three B-sites (see Fig.~\ref{fig:hex_unit}(a)), and an enlarged sixteen sites unit cell for the $\bm{M}_{n}$ instability, we numerically minimize the total energy and obtain the ground-state spin configuration. For the sixteen site unit-cell calculation, we include only $U$-terms of the interactions (\ref{eq:heff_2mag}).

\subsubsection{Phase I: Coplanar stripes phase}

%*%*%*%*%*%*%*%*%*%*%*%*%*%*%*%*%*%*%*
%  fig8
%*%*%*%*%*%*%*%*%*%*%*%*%*%*%*%*%*%*%*
\begin{figure}
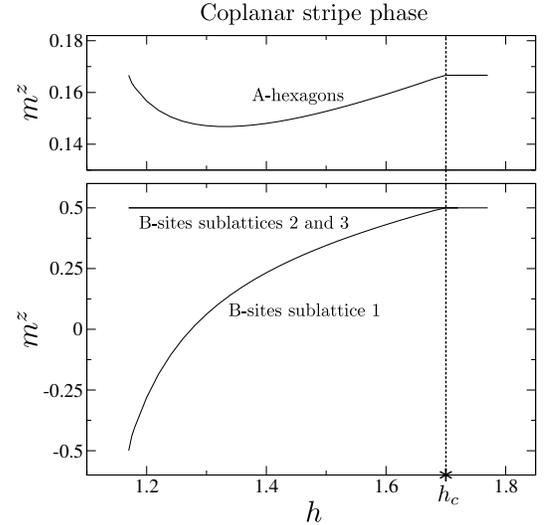

\begin{center}
\includegraphics[width=0.8\columnwidth,clip]{{{submag_J_084}}}
\end{center}
\caption{Local magnetizations in the supersolid phase at $1/3 < m \le 5/9$ for $J'=0.84$. These values are evaluated from the MF approximation of the effective Hamiltonian (\ref{eq:H_eff_int}). Top: Magnetization per site in each resonating hexagon. Bottom: Sublattice magnetizations of the three B-site sublattices. The star symbol indicates the plateau lower critical field $h_{c} = 1.70$.}
\label{fig:submag_J.0.84}
\end{figure}
%*%*%*%*%*%*%*%*%*%*%*%*%*%*%*%*%*%*%*
%*%*%*%*%*%*%*%*%*%*%*%*%*%*%*%*%*%*%*

At $J^\prime/J=0.84$, we have seen in Sec.~\ref{sec:magnon_insta} that the one-magnon instability appears at the three wave vectors ${\bm M}_n$. Therefore, we assume a spin structure with a four times larger unit cell to allow for a possible superposition of magnon condensates at different wave vectors ${\bm M}_{n}$.

We find that the ground state is in fact not a superposition but a coplanar stripe state characterized by only one of the ${\bm M}_n$. More precisely, magnons on the sublattice $\Lambda_n$ and A-hexagons form a Bose-Einstein condensation (BEC) at one selected wave vector $\bm{M}_n$, giving a finite expectation value $\langle d^\dagger_{n,{\bm M}_n} \rangle \propto \sqrt{N}e^{i\varphi}$ along stripes in real space. For instance, the spin configuration corresponding to the $\bm{M}_1$ wave vector can be written as
\begin{align}
  \langle {\bm T}_I \rangle & =\frac{1}{2}(\exp(i{\bm M}_1\cdot {\bm r}_I)\sin \theta_0,0,\cos \theta_0),\nonumber\\
  \langle {\bm S}_i \rangle & =
    \frac{1}{2} (-\exp(i{\bm M}_1\cdot {\bm r}_i) \sin \theta_1,0,\cos \theta_1) &  \ \  (i\in\Lambda_1),
  \label{eq:spins_phase1}
\end{align}
whereas the spins on the B-site sublattices $\Lambda_2$ and $\Lambda_3$ remain fully polarized. Here, the origin of the coordinates is located on a hexagon center, and the canting angles, $0\le\theta_0\le \pi$ and $0\le\theta_1\le \pi$, are numerically evaluated. Because of the non-zero wave vector, the unit cell is doubled in the direction ${\bm r}\parallel {\bm M}_1$ and contains eight spins, \textit{i.e.} two A-hexagons and six B-sites. Thus, the supersolid phase breaks the $C_3$ space rotation symmetry by choosing one of the three wave vectors. In total, the transition from the $m=5/9$ plateau to the supersolid phase at lower magnetizations is accompanied by the $U(1)\times C_3$ symmetry breaking. The local magnetizations in the original kagome lattice are numerically evaluated in Fig.~\ref{fig:submag_J.0.84}. One sees that even though we introduced magnon interactions, the B-sites on sublattices $\Lambda_2$, $\Lambda_3$, remain fully polarized, and the magnetization slope comes only from the A-hexagons and $\Lambda_1$ sublattice.

The physical picture of the supersolid phase is thus the alternative alignment of stripes formed by the superfluid and supersolid components. The rigid solid component is protected by the kinetic frustration effect. As shown in Fig.~\ref{fig:supersolid_configs}(a), the $\Lambda_2$ is connected to the two A-hexagons on different stripes, and since A-hexagons have an alternating phase, the contributions of hopping to $\Lambda_2$ B-sites from the two superfluid stripes cancel out. This happens also for $\Lambda_3$ B-sites. We conclude by reminding that although the curve presented in Fig.~\ref{fig:submag_J.0.84} extends down to $m=1/3$, the validity of our approach breaks down before reaching the $1/3$-plateau and another phase transition might occur.

\subsubsection{Phase II: Coplanar V phase}

%*%*%*%*%*%*%*%*%*%*%*%*%*%*%*%*%*%*%*
%  fig9
%*%*%*%*%*%*%*%*%*%*%*%*%*%*%*%*%*%*%*
\begin{figure}
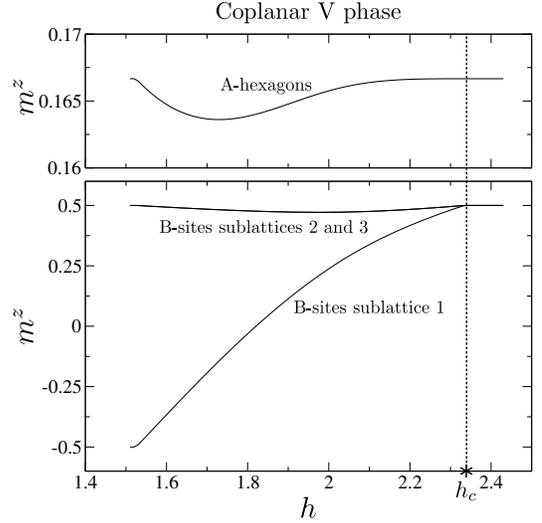

\begin{center}
\includegraphics[width=0.8\columnwidth,clip]{{{submag_J_1}}}
\end{center}
\caption{Local magnetizations in the supersolid phase at $1/3 < m \le 5/9$ for the kagome antiferromagnet $J'=1$. These values are calculated from the MF approximation of the effective Hamiltonian (\ref{eq:H_eff_int}). Top: Magnetization per site in each resonating hexagon. Bottom: Sublattice magnetizations of the three B-sites sublattices. The star symbol indicates the plateau lower critical field $h_{c}=2.33$.}
\label{fig:submag_J.1}
\end{figure}
%*%*%*%*%*%*%*%*%*%*%*%*%*%*%*%*%*%*%*
%*%*%*%*%*%*%*%*%*%*%*%*%*%*%*%*%*%*%*

In the case of $J^{\prime}=1$, we have shown in Sec.~\ref{sec:magnon_insta} that the instability appears at the $\Gamma$ point, with
a degeneracy originating from a chiral degree of freedom defined on the B-sites. In our MF calculation, the numerically obtained spin structure preserves the four sites unit cell of the Hamiltonian with one hexagon and three B-sites, consistent with the wave vector ${\bm \Gamma}$. To confirm our finding, we performed the energy minimization for the extended sixteen sites unit cells in the whole magnetization range $1/3 < m \le 5/9$.

\begin{figure}
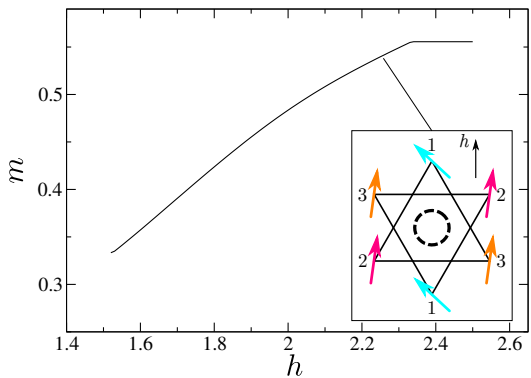

\begin{center}
\includegraphics[width=0.8\columnwidth,clip]{{{mag_J_1}}}
\end{center}
\caption{(Color online) Magnetization per spin $m$ between the two plateaus at $m=1/3$ and $5/9$ for $J^{\prime}=1$,
as evaluated from the MF approximation of the effective Hamiltonian (\ref{eq:H_eff_int}). Inset: representation of the supersolid spin structure (V state) around a resonating hexagon.}
\label{fig:mag_J.1}
\end{figure}

In the ground state, we find that the spins on B-sites have the three-sublattice structure of the coplanar V state configuration shown in Fig.~\ref{fig:supersolid_configs}. In Fig.~\ref{fig:submag_J.1}, we plot the sublattices magnetizations and see that this V structure has the feature that two sublattices are still almost fully polarized. Below the plateau, the A-hexagons keep the original magnetization $S^z_{\hexagon}=1$ coming from the $|1,0\rangle_A$ states, and the spin structure on the B-sites is given by the superposition of the two condensates $\langle d_{+,{\bm \Gamma}}^\dagger \rangle =\langle d_{-,{\bm \Gamma}}^\dagger \rangle \propto \sqrt{N}e^{i\varphi}$ with an equal weight and an equal phase. Further decreasing the magnetic field, the spins on the A-hexagons gradually start to cant. One of the spin configurations can be written as
\begin{align}
  \langle {\bm T}_I \rangle & =\frac{1}{2}( \sin \theta_0,0,\cos \theta_0) ,\nonumber\\
  \langle {\bm S}_i \rangle & =
  \left\{ \begin{array}{ll}
    \frac{1}{2} (- \sin \theta_1,0,\cos \theta_1) &  \ \ \ (i \in\Lambda_1),\\
    \frac{1}{2}(\sin \theta_2,0,\cos \theta_2) &  \ \ \ (\text{$i \in\Lambda_2, \Lambda_3$}),
  \end{array} \right.
  \label{eq:spins_phase2}
\end{align}
with the numerically evaluated $0\le\theta_n \le \pi$, $(n=0,1,2)$. Using the relation (\ref{eq:mag_h}) between the hexagon pseudo-spin and the original A-sites spins, we see that this translates into a coplanar configuration over the whole lattice. The magnetization per spin $m$ of the whole lattice, given by
\begin{equation}\label{eq:m}
  m =  \frac{1}{9}(1+\sin \theta_0 - \sin \theta_1 + 2 \sin \theta_2),
\end{equation}
is plotted in Fig.~\ref{fig:mag_J.1}. The curve shows a finite slope comparable to the numerical results.~\cite{Nishimoto2013}

%*%*%*%*%*%*%*%*%*%*%*%*%*%*%*%*%*%*%*
%  fig10
%*%*%*%*%*%*%*%*%*%*%*%*%*%*%*%*%*%*%*
%*%*%*%*%*%*%*%*%*%*%*%*%*%*%*%*%*%*%*
%*%*%*%*%*%*%*%*%*%*%*%*%*%*%*%*%*%*%*

The state below $m=5/9$ developing from the $\Gamma$ point instability is therefore a supersolid whose
solid component keeps the resonating hexagonal structure which has the same symmetry as that of the $m=5/9$ plateau ($\sqrt{3}\times\sqrt{3}$) state,
whereas the transverse (superfluid) component breaks the U(1) spin rotation symmetry and space $C_3$ rotation symmetries. The transition out of the $m=5/9$ plateau state to the supersolid is therefore accompanied by the $U(1) \times C_3$ symmetry breaking.

Comparing Fig.~\ref{fig:submag_J.0.84} and Fig.~\ref{fig:submag_J.1}, it is interesting to remark that the magnetization curves below the plateau are similar between the two coplanar stripes and V phases. Namely, one of the B-site sublattices carries the magnetization process, assisted by the A-hexagons.
The two other B-site sublattices are either fully polarized or very weakly canted. The symmetries of the two phases are nonetheless different.
Thus, even though we cannot predict for sure that the V phase is realized for the kagome limit $J^{\prime}=1$ rather than the stripes phase,
this common property of which subset of the sites participate in the magnetization process
(or, equivalently, to the superfluid) could be an intrinsic feature of a supersolid below the $m=5/9$ plateau.

%%%%%%%%%%%%%%%%%%%%%%%%%%%%%%%%%%%%%%%%%%%%%%%%%%%%%%%%%%%%%%%%%
%%%%%%%%%%%%%%%%%%%%%%%%%%%%%%%%%%%%%%%%%%%%%%%%%%%%%%%%%%%%%%%%%
\section{Other plateaus}
\label{sec:other_plateaus}
%%%%%%%%%%%%%%%%%%%%%%%%%%%%%%%%%%%%%%%%%%%%%%%%%%%%%%%%%%%%%%%%%
%%%%%%%%%%%%%%%%%%%%%%%%%%%%%%%%%%%%%%%%%%%%%%%%%%%%%%%%%%%%%%%%%

Finally, we briefly report our results on the magnon instabilities for the plateaus at $m=1/3$ and 7/9, and for the upper boundary of the 5/9 plateau.
To analyze them, we use essentially the same method as in Sec.~\ref{sec:perturb}.
The only difference is in the choice of the low-energy degenerate space, which depends on the energy level structure of Fig.~\ref{fig:hex_unit}.

%%%%%%%%%%%%%%%%%%%%%%%%%%%%%%%%%%%%%%%%%%%%%%%%%%%%%%%%%%%%%%%%%
\subsection{$m=1/3$ plateau}
%%%%%%%%%%%%%%%%%%%%%%%%%%%%%%%%%%%%%%%%%%%%%%%%%%%%%%%%%%%%%%%%%

One may expect that the lower magnetization regime in the range $1/3 < m < 5/9$ is accessible from the higher critical field of the 1/3 plateau phase. This is, however, not the case. To see it, we start from the exact ground state at $J^{\prime}=0$,
\begin{align}
|\Psi_{1/3}\rangle=  \prod_I |0,0 \rangle_{A;I} \prod_{i} |\uparrow \rangle_{B;i},
\label{eq:m1over3}
\end{align}
and calculate the one-magnon excitation energy in the sector $\Delta S^z=+1$.

The lowest magnon state from $|\Psi_{1/3}\rangle$ is created by exciting an A-hexagon to $|1,0\rangle_{A;I}$.
For the same symmetry reason as in Sec.~\ref{sec:magnon_insta}, the magnon is localized at any order in $J^{\prime}$. Up to third order, the higher critical field is $h_{c} = 0.685 + J^{\prime} + 0.274 J^{\prime 2} - 0.17 J^{\prime 3}$, which takes the value $h_{c}=1.79$ at $J^{\prime}=1$.
This is quite far from the DMRG value $h_{c}^{\mathrm{DMRG}} \simeq 1.2$.~\cite{Nishimoto2013}
We further considered the second and third one-magnon states, obtained by exciting a hexagon to the second and third hexagon states $|1,1\rangle_{A;I}$ and $|1,2\rangle_{A;I}$ respectively, which are both two-fold degenerate.
We find that those four magnons are dispersive but all have very weak hoppings amplitudes. Consequently, the aforementioned flat band always remains the lowest energy excitation. The fourth magnon state is also localized and remains a high energy excitation.

Therefore, none of the magnon excitations studied appears to be relevant for the transition at the higher critical field of the plateau $m=1/3$.
This suggests that the transition would rather be of first order and hence not accessible from our method. This is consistent with the 36 sites exact diagonalization spectrum, which found no evidence of the crystal order just above the plateau.~\cite{Capponi2013}

%%%%%%%%%%%%%%%%%%%%%%%%%%%%%%%%%%%%%%%%%%%%%%%%%%%%%%%%%%%%%%%%%
\subsection{Upper boundary of $m=5/9$ plateau}
%%%%%%%%%%%%%%%%%%%%%%%%%%%%%%%%%%%%%%%%%%%%%%%%%%%%%%%%%%%%%%%%%

The lowest one-magnon state in the sector $\Delta S^z=+1$ starting from $|\Psi_{5/9}\rangle$ has one excited $|2,0\rangle_{A;I}$ hexagon. Like in the other cases, its $k_{\hexagon}=\pi$ momentum results in an exactly localized magnon. At third order and for $J^{\prime}=1$, we evaluate the upper critical field $h_c \sim 2.68$, which is comparable to the DMRG value.~\cite{Nishimoto2013} Two degenerate magnon states are then obtained by having one hexagon in the twofold degenerate $|2,1\rangle_{A;I}$ state. Up to third order, those magnons have a small chemical potential, as well as negligible hopping amplitudes ($\sim 10^{-3}$ at $J^{\prime}=1$).
Therefore, they cannot overcome the flat mode.
We also note that, combined with the results from Sec.~\ref{sec:magnon_insta},
the $m=5/9$ plateau is stable against magnon excitations in the whole range of $J'$, particularly at $J'=1$ in a finite field range $2.3 \le h \le 2.68$.
We find the same almost dispersionless behaviour for the third magnon state. According to the DMRG results~\cite{Nishimoto2013} (but not to the iPEPS results~\cite{Picot2016}), the magnetization process at the plateau upper boundary is very steep. This could suggest that some nearly flat modes contribute to the transition. However, it is beyond our scope to deal with magnon instabilities that have magnetization differences from the plateau value by more than one and thus we conclude by stating that our finding of magnons with extremely small bandwidths can be compatible with numerics.

%%%%%%%%%%%%%%%%%%%%%%%%%%%%%%%%%%%%%%%%%%%%%%%%%%%%%%%%%%%%%%%%%
\subsection{$m=7/9$ plateau}
%%%%%%%%%%%%%%%%%%%%%%%%%%%%%%%%%%%%%%%%%%%%%%%%%%%%%%%%%%%%%%%%%

Starting from $|\Psi_{7/9}\rangle=\prod_{I}|2,0\rangle_{A;I} \prod_{i} |\uparrow \rangle_{B;i}$ (which is the exact plateau ground state also at $J^{\prime}=1$~\cite{}), we excite one hexagon to $|1,0\rangle_{A;I}$ to construct the lowest one-magnon state.
As for the previous lowest magnon states, the effective hopping exactly vanishes. The magnon is thus localized and cannot explain the transition. Two degenerate second magnon states are created by promoting one hexagon to one of the twofold degenerate hexagon states $|1,1\rangle_{A;I}$.
They are, however, difficult to analyze within our perturbative scheme.
Indeed, we find that the prefactors of the chemical potentials and the hoppings are almost equal for the second and third order contributions.
This indicates that the perturbation is not converged at all;
thus we cannot reliably predict if those magnons can achieve a lower energy than the flat mode.
The similar lack of convergence is found in the third magnon excitation, which is also twofold degenerate.

Finally, we remind that the transition at the upper critical field, namely the magnetization jump between the plateau and the saturated state, is already known to be the condensation of the exactly localized $|2,0\rangle_{A;I}$ magnons.~\cite{}

%%%%%%%%%%%%%%%%%%%%%%%%%%%%%%%%%%%%%%%%%%%%%%%%%%%%%%%%%%%%%%%%%
%%%%%%%%%%%%%%%%%%%%%%%%%%%%%%%%%%%%%%%%%%%%%%%%%%%%%%%%%%%%%%%%%
\section{Conclusion}
\label{sec:conclusion}
%%%%%%%%%%%%%%%%%%%%%%%%%%%%%%%%%%%%%%%%%%%%%%%%%%%%%%%%%%%%%%%%%
%%%%%%%%%%%%%%%%%%%%%%%%%%%%%%%%%%%%%%%%%%%%%%%%%%%%%%%%%%%%%%%%%

The magnetization process of the spin-1/2 kagome antiferromagnet is known to be rich,
including four plateaus at $m=1/9,3/9,5/9,7/9$.~\cite{Nishimoto2013,Capponi2013} In this paper, we disclosed that
{\it the kinetic frustration effect} plays a key role to protect the plateau phases against the instability toward forming a superfluid,
and generate a supersolid phase just below the $m=5/9$ phase.

On the plateaux $m=1/3$, 5/9 and 7/9, numerical simulations revealed a magnetic nine site unit cell, further decomposed into a ``resonating" hexagon and three fully polarized neighbouring spins. We studied a $J-J^{\prime}$ model on a hexamerized lattice, explicitly partitioning the lattice between the hexagons (A-hexagons) and the polarized spins (B-sites). We then examined the possible instabilities of the plateaux upon varying the magnetic field, by perturbatively calculating the energies of one-magnon excitations from the exact limit $J^{\prime}=0$ towards the kagome lattice $J^{\prime}=J$.
Within our approach, the condensation of magnons formally corresponds to the formation of a supersolid phase.

We particularly focused our analysis on the instabilities at the lower boundary of the $m=5/9$ plateau.
The points clarified are summarized as follows:
The excitation from the plateau is described by the introduction of a magnon on each A-hexagon and/or B-site,
and the symmetry of the Wannier wave functions of magnons on A-hexagons turned out to be important.
Many of the magnons remain localized on A-hexagons, since the symmetry of their wave functions
allows the destructive interference of two hopping paths from the adjacent sites on A-hexagons to the neighboring B-sites,
which is a typical {\it kinetic frustration} effect.
Resultantly, only part of the A-magnons become really dispersive, and contribute to the normal one-magnon instability.

This is, however, not the end of the story.
Through virtual perturbation processes mediated by the A-hexagons,
B-magnons acquire direct itinerancy to further B-site-neighbours,
a hopping not explicitly present in the original model.
These couplings eventually become relatively strong.
The resultant hopping paths are complicated and competing, which work destructively with each other,
generating several minima in their energy dispersion.
A large difference in chemical potential between A and B-magnons also induced by the perturbation
works to select part of these competing hopping paths, which effectively works as another aspect of kinetic frustration.
Eventually, by introducing an inter-magnon interaction at the same order of perturbation
and in recovering the original model parameters of the uniform kagome antiferromagnet at $J'=J$,
we reach a supersolid phase where one of the three B-sublattices dominate the magnetization process,
while the A-hexagons keep the resonating structure of the $m=5/9$ plateau.

The instabilities at other plateau boundaries are also studied within the same perturbative approach. For all the transitions we considered, we essentially found that none can be reliably described by a simple one-magnon condensation. Indeed, the lowest one-magnon state always gives an exact flat band, and higher excitations do not achieve a lower energy, mostly due to very weak effective hoppings. It suggests that we should look for other mechanisms in order to describe the transitions other than below $m=5/9$.

The supersolid discussed in the atomic Bose gas systems and in the XXZ quantum spin systems were supported by the extremely large competing interaction
with the aid of small quantum fluctuation.
In the present system, however, an applied magnetic field partially resolves the degeneracies
of the energies, inducing crystal structure of resonating hexagons.
The presence of resonating hexagons efficiently suppresses
the kinetic energy scale as well as select the path of superfluid in real space,
providing good description of how such supersolids may further enrich
this celebrated magnetization curve.

%%%%%%%%%%%%%%%%%%%%%%%%%%%%%%%%%%%%%%%%%%%%%%%%%%%%%%%%%%%%%%%%%
%%%%%%%%%%%%%%%%%%%%%%%%%%%%%%%%%%%%%%%%%%%%%%%%%%%%%%%%%%%%%%%%%

\begin{acknowledgments}
We acknowledge useful discussions with S.~Capponi and A.~Furusaki. This work was supported by JSPS KAKENHI Grant Numbers JP16K05425, JP17K05533,
JP17K05497, and JP17H02916. X.P. was supported by RIKEN iTHES Project.
\end{acknowledgments}

%%%%%%%%%%%%%%%%%%%%%%%%%%%%%%%%%%%%%%%%%%%%%%%%%%%%%%%%%%%%%%%%%
%%%%%%%%%%%%%%%%%%%%%%%%%% APPENDIX %%%%%%%%%%%%%%%%%%%%%%%%%%%%%
%%%%%%%%%%%%%%%%%%%%%%%%%%%%%%%%%%%%%%%%%%%%%%%%%%%%%%%%%%%%%%%%%

\appendix

%%%%%%%%%%%%%%%%%%%%%%%%%%%%%%%%%%%%%%%%%%%%%%%%%%%%%%%%%%%%%%%%%
%%%%%%%%%%%%%%%%%%%%%%%%%%%%%%%%%%%%%%%%%%%%%%%%%%%%%%%%%%%%%%%%%
\section{Derivation of the effective Hamiltonians}
\label{app:perturb}
%%%%%%%%%%%%%%%%%%%%%%%%%%%%%%%%%%%%%%%%%%%%%%%%%%%%%%%%%%%%%%%%%
%%%%%%%%%%%%%%%%%%%%%%%%%%%%%%%%%%%%%%%%%%%%%%%%%%%%%%%%%%%%%%%%%

We briefly outline the derivation of the effective Hamiltonians Eq.(\ref{eq:heff_1mag_loc}), Eq.(\ref{eq:heff_1mag}) and Eqs.(\ref{eq:heff_2mag})-(\ref{eq:heff_2mag_proc}), within the framework of degenerate perturbation theory.

%%%%%%%%%%%%%%%%%%%%%%%%%%%%%%%%%%%%%%%%%%%%%%%%%%%%%%%%%%%%%%%%%
\subsection{Degenerate perturbation theory}
\label{app:perturb_general}
%%%%%%%%%%%%%%%%%%%%%%%%%%%%%%%%%%%%%%%%%%%%%%%%%%%%%%%%%%%%%%%%%

We use the standard formalism proposed by Bloch.~\cite{Bloch1958} Given an unperturbed Hamiltonian $\mathcal{H}_{0}$, we define $P_0$ and $Q_0$ the projection operators onto the degenerate unperturbed ground state manifold of energy $E_0$ and its complement, respectively. For a perturbation $\lambda V$, the effective Hamiltonian up to third order reads
\begin{equation}
\begin{split}
\mathcal{H}_{\mathrm{eff}} & = E_0 + \lambda P_0 V P_0 + \lambda^2 P_0 V \frac{1}{E_0-\mathcal{H}_0} Q_0 V P_{0} \\
& + \lambda^{3} \left[ P_0 V \frac{1}{E_0-\mathcal{H}_0} Q_0 V \frac{1}{E_0-\mathcal{H}_0} Q_0 V P_{0} \right. \\
& \left. - P_0 V \frac{1}{(E_0-\mathcal{H}_0)^{2}} Q_0 V P_0 V P_0 \right].
\end{split}
\end{equation}
Because of the last term, the effective Hamiltonian is in general non-hermitian at third order. In such case, we have taken the hermitian combination $(\mathcal{H}_{\mathrm{eff}}+\mathcal{H}_{\mathrm{eff}}^{\dagger})/2$.

%%%%%%%%%%%%%%%%%%%%%%%%%%%%%%%%%%%%%%%%%%%%%%%%%%%%%%%%%%%%%%%%%
\subsection{Perturbative processes and Hamiltonians parameters}
\label{app:perturb_example}
%%%%%%%%%%%%%%%%%%%%%%%%%%%%%%%%%%%%%%%%%%%%%%%%%%%%%%%%%%%%%%%%%

%*%*%*%*%*%*%*%*%*%*%*%*%*%*%*%*%*%*%*
%  fig11
%*%*%*%*%*%*%*%*%*%*%*%*%*%*%*%*%*%*%*
\begin{figure}
\begin{center}
\includegraphics[width=0.98\columnwidth,clip]{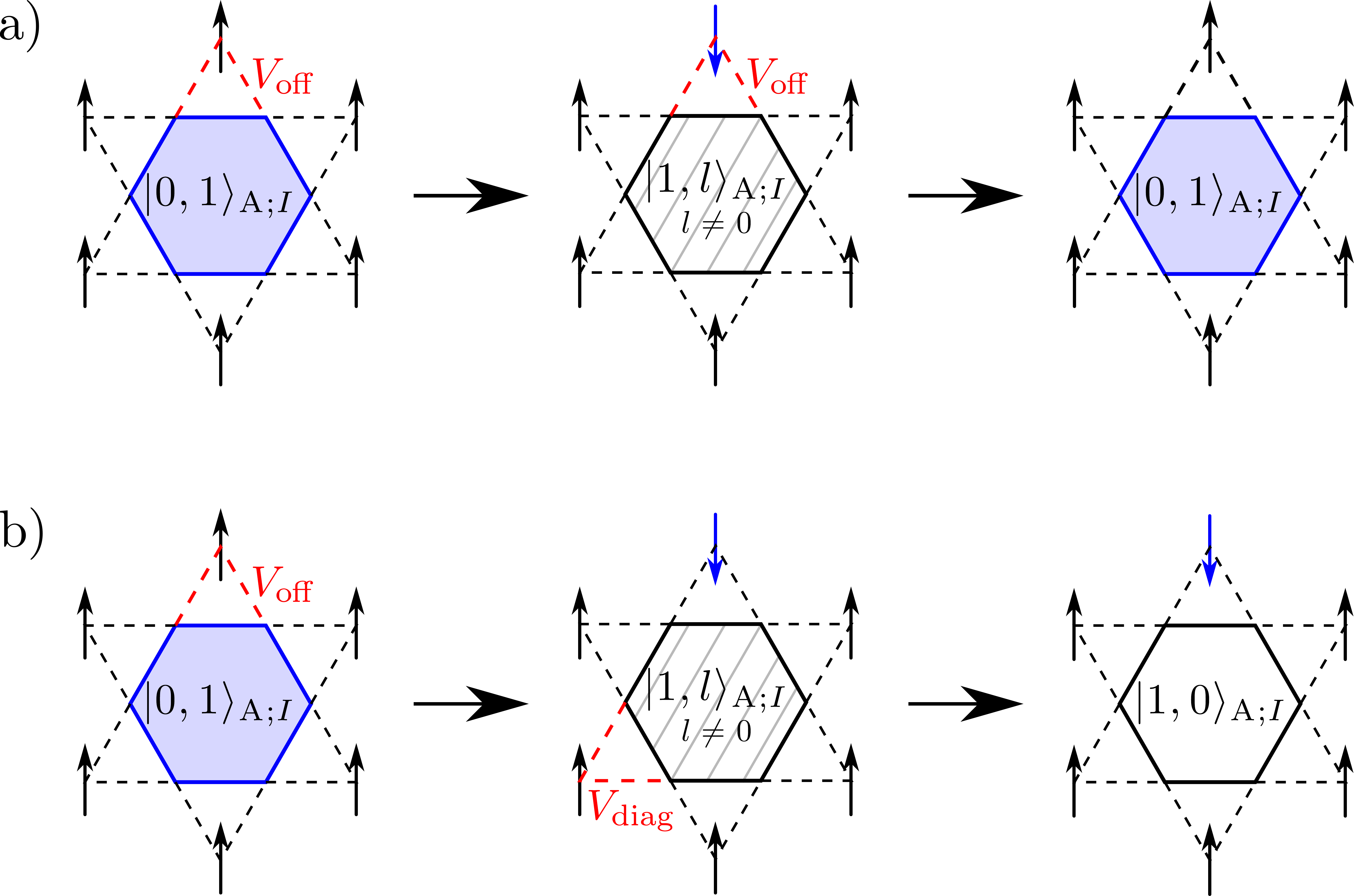}
\end{center}
\caption{(Color online) (a) Diagonal second order perturbative process on a $|0,1\rangle_{\mathrm{A};I}$ hexagon, or A-hexagon magnon, which contributes to the chemical potential $\mu_{A}$. Blue color indicates the presence of a magnon on the A-hexagon or a B-site, and striped filling excited intermediate states. The perturbation is separated between its diagonal and off-diagonal terms $V_{\mathrm{diag}}$ and $V_{\mathrm{off}}$. (b) Off-diagonal second order process responsible for the hopping $t_{AB;1}$.}
\label{fig:app_perturb}
\end{figure}
%*%*%*%*%*%*%*%*%*%*%*%*%*%*%*%*%*%*%*
%*%*%*%*%*%*%*%*%*%*%*%*%*%*%*%*%*%*%*

The derived effective Hamiltonians describe the one-magnon excitations above the plateau state. After creating a magnon on either an A-site (hexagon) or a B-site (polarized spins), we rewrite the diagonal terms of the effective Hamiltonian as
\begin{equation}
\langle \Psi^{\mathrm{ex}} |\mathcal{H}_{\mathrm{eff}} | \Psi^{\mathrm{ex}} \rangle = E_{5/9} - (\mu_{\mathrm{ex}} + h),
\end{equation}
which defines the chemical potential $\mu_{\mathrm{ex}}$ corresponding to an excitation on either the A-hexagons or the B-sites. In this expression, $| \Psi^{\mathrm{ex}} \rangle$ is one of the magnon states from (\ref{eq:1st_exc}) or (\ref{eq:2nd_exc}), and $E_{5/9}$ is the energy of the plateau state at the same order in perturbation. In the Hamiltonians reported in the main text, we have dropped the constant $E_{5/9}$ such that the excitation energy is zero at the plateau transition. Similarly, the excited states with neighbouring magnons produce diagonal terms including both the chemical potentials and the two-magnon interaction terms.

Practically, the chemical potential essentially corresponds to the difference between the diagonal processes on a plateau state $|1,0\rangle$ and on an excited hexagon or flipped spin. For instance, the second order contribution to $\mu_{A}$ is
\begin{equation}
\begin{split}
\mu_{A}^{(2)} = J'^{2} \sum_{n=1}^{6} & \left( \sum_{l} \frac{|\langle 2,l |S^{+}_{n}|1,0\rangle|^{2}}{E_{0}-E_{2,l}} \right. \\
& \left. - \sum_{l \neq 0} \frac{|\langle 1,l |S^{+}_{n}|0,1\rangle|^{2}}{E_{0}-E_{1,l}} \right),
\end{split}
\end{equation}
where $E_{0} = E_{0,1}=E_{1,0}$ and $n$ labels the six sites of a A-hexagon. The diagonal process corresponding to the second term is represented in Fig.~\ref{fig:app_perturb}(a).

Similar procedure gives the hopping parameters of the effective Hamiltonians. In Fig.~\ref{fig:app_perturb}(b), we illustrate the off-diagonal process which contributes at second order to the hopping $t_{AB;1} \equiv {}_{i}\langle \Psi^{\rm ex}_{b} |\mathcal{H}_{\mathrm{eff}} | \Psi^{\rm ex}_{a} \rangle_I$, with $i$ and $I$ nearest neighbours. After simplification, we obtain
\begin{equation}
t_{AB;1}^{(2)} = -J'^{2} \sum_{n,m=1}^{2} \sum_{l \neq 0} \frac{\langle 1,0 |S^{z}_{n}|1,l\rangle \langle 1,l |S^{+}_{m}|0,1\rangle}{E_{0}-E_{1,\alpha}}.
\end{equation}
At third order, an example of contributing process is obtained by inserting an additional diagonal operator in second position, which brings the hexagon to another intermediate state $|1,l'\rangle$.

%%%%%%%%%%%%%%%%%%%%%%%%%%%%%%%%%%%%%%%%%%%%%%%%%%%%%%%%%%%%%%%%%
%%%%%%%%%%%%%%%%%%%%%%%%%%%%%%%%%%%%%%%%%%%%%%%%%%%%%%%%%%%%%%%%%
\section{Parameters of the effective Hamiltonians}
\label{app:params}
%%%%%%%%%%%%%%%%%%%%%%%%%%%%%%%%%%%%%%%%%%%%%%%%%%%%%%%%%%%%%%%%%
%%%%%%%%%%%%%%%%%%%%%%%%%%%%%%%%%%%%%%%%%%%%%%%%%%%%%%%%%%%%%%%%%

In this appendix, we report the numerical values of the parameters in the effective Hamiltonians (\ref{eq:heff_1mag}) and (\ref{eq:heff_2mag}), up to order of $J'^{3}$.

%%%%%%%%%%%%%%%%%%%%%%%%%%%%%%%%%%%%%%%%%%%%%%%%%%%%%%%%%%%%%%%%%
\subsection{One-magnon effective Hamiltonian}
\label{app:heff_1mag_param}
%%%%%%%%%%%%%%%%%%%%%%%%%%%%%%%%%%%%%%%%%%%%%%%%%%%%%%%%%%%%%%%%%

The hopping amplitudes of the one-magnon effective Hamiltonian ${\cal H}_{\rm kin}$ [Eq.~(\ref{eq:heff_1mag})]
in the sector $\Delta S^z=-1$ are given by
\begin{equation}
\begin{split}
t_{AB,1} &= 0.236 J' + 0.151 J'^{2} - 0.135 J'^{3} ,\\
t_{AB,3} &= 0.0003 J'^{3} ,\\
t_{AB,3} &= 0.007 J'^{3} ,\\
t_{AB,4} &= -0.0005 J'^{3} ,\\
t_{BB,1} &= 0.0006 J'^{2} + 0.052 J'^{3},\\
t_{BB,2} &= 0.07 J'^{2} + 0.141 J'^{3},\\
t_{BB,3} &= -0.001 J'^{2} + 0.021 J'^{3},\\
t_{AA} &= 0.038 J'^{3},
\end{split}
\end{equation}
and the chemical potentials by
\begin{equation}
\begin{split}
\mu_{A} &=  J' + 0.639 J'^{2} - 0.283 J'^{3},\\
\mu_{B} &= \frac{2}{3} J' + 0.705 J'^{2} + 0.473 J'^{3},
\end{split}
\end{equation}

%%%%%%%%%%%%%%%%%%%%%%%%%%%%%%%%%%%%%%%%%%%%%%%%%%%%%%%%%%%%%%%%%
\subsection{Two-magnon effective interactions}
\label{app:heff_2mag_param}
%%%%%%%%%%%%%%%%%%%%%%%%%%%%%%%%%%%%%%%%%%%%%%%%%%%%%%%%%%%%%%%%%

The coupling constants of two-magnon interactions ${\cal V}_{\rm int}$ [Eq.~(\ref{eq:heff_2mag})] are obtained as follows:
The assisted hopping processes have amplitudes
\begin{equation}
\begin{split}
T^{AB;B}_{1} &= -0.008 J'^{2} + 0.006 J'^{3}, \\
T^{AB;B}_{2} &= -0.069 J'^{2} - 0.045 J'^{3}, \\
T^{AB;B}_{3} &= 0.017 J'^{2} + 0.022 J'^{3}, \\
T^{BB;A}_{1} &= 0.006 J'^{2} + 0.014 J'^{3},\\
T^{BB;A}_{2} &= 0.049 J'^{2} - 0.040 J'^{3},\\
T^{BB;A}_{3} &= -0.012 J'^{2} - 0.021 J'^{3},\\
T^{BB;B}_{1,2,1} &= 0.055 J'^{3},\\
T^{BB;B}_{1,2,3} &= -0.011 J'^{3},\\
T^{BB;B}_{2,1,3} &= 0.015 J'^{3},\\
T^{BB;B}_{2,2,2} &= -0.053 J'^{3},\\
T^{BB;B}_{3,1,2} &= -0.002 J'^{3},
\end{split}
\end{equation}
where the hermiticity imposes that the $B_{cs}$ parameters satisfy $T^{BB;B}_{r_1,r_2,r_3} = T^{BB;B}_{r_1,r_3,r_2}$.
There are six distinct non-zero pair hoppings,
\begin{equation}
\begin{split}
\gamma_{1,1,2} &= 0.039 J'^{3},\\
\gamma_{1,2,3} &= -0.01 J'^{3},\\
\gamma_{2,1,1} &= -0.07 J'^{3},\\
\gamma_{2,1,3} &= 0.014 J'^{3},\\
\gamma_{2,2,2} &= 0.094 J'^{3},\\
\gamma_{3,1,2} &= -0.008 J'^{3}.
\end{split}
\end{equation}
The density-density interactions, which contain the dominant two-magnon processes, have the values
\begin{equation}
\begin{split}
U_{AB} &= \frac{1}{3} J' + 0.311 J'^{2} - 0.155 J'^{3},\\
U_{BB,1} &= 0.013 J'^{2} + 0.139 J'^{3},\\
U_{BB,2} &= 0.237 J'^{2} + 0.401 J'^{3},\\
U_{BB,3} &= -0.026 J'^{2} - 0.014 J'^{3},\\
U_{AA} &= 0.076 J'^{3}.
\end{split}
\end{equation}

%%%%%%%%%%%%%%%%%%%%%%%%%%%%%%%%%%%%%%%%%%%%%%%%%%%%%%%%%%%%%%%%%
%%%%%%%%%%%%%%%%%%%%%%%%%%%%%%%%%%%%%%%%%%%%%%%%%%%%%%%%%%%%%%%%%
\section{Effective spin Hamiltonian}
\label{app:mf_ham}
%%%%%%%%%%%%%%%%%%%%%%%%%%%%%%%%%%%%%%%%%%%%%%%%%%%%%%%%%%%%%%%%%
%%%%%%%%%%%%%%%%%%%%%%%%%%%%%%%%%%%%%%%%%%%%%%%%%%%%%%%%%%%%%%%%%

Using the pseudo-spin operators defined by $T^z_I:=1/2 -a^{\dagger}_{I} a_{I}$ and $T^{-}_{I}:=a^{\dagger}_{I}$,
and also using the original spin operators on B-sites, the effective  Hamiltonian (\ref{eq:H_eff_int}) in the spin representation reads
\begin{align}
\mathcal{H}_{\rm eff} &= \sum_{\langle iJ \rangle} J^{xy}_{AB} ( S^{x}_{i} T^{x}_{J} + S^{y}_{i} T^{y}_{J} ) + J^{z}_{AB} S^{z}_{i} T^{z}_{J} \nonumber\\
&+ \sum_{r=1}^{3} \sum_{\langle ij \rangle_{r}} J^{xy}_{BB,r} ( S^{x}_{i} S^{x}_{j} + S^{y}_{i} S^{y}_{j} ) + J^{z}_{BB,r} S^{z}_{i} S^{z}_{j} \nonumber\\
&+ \sum_{\langle IJ \rangle} J^{xy}_{AA} ( T^{x}_{I} T^{x}_{J} + T^{y}_{I} T^{y}_{J} ) + J^{z}_{AA} T^{z}_{I} T^{z}_{J} \nonumber\\
&- 2 \sum_{I}  \sum_{r=1}^{3} \sum_{\langle jk \rangle_r \in\hexagon_I} \{ T^{AB;B}_r (S^{x}_{j} T^{x}_{I} + S^{y}_{j} T^{y}_{I} ) S^{z}_{k} \nonumber\\
&%- 2 \sum_{I}  \sum_{r=1}^{3} \sum_{\langle jk \rangle_r \in\hexagon_I}
\qquad \qquad \qquad \qquad +T^{BB;A}_{r} (S^{x}_{j} S^{x}_{k} + S^{y}_{j} S^{y}_{k}) T^{z}_{I} \} \nonumber\\
&- 2 \sum_{I} \sum_{jkl\in\hexagon_{I}} T^{BB;B}_{r_1,r_2,r_3} ( S^{x}_{j} S^{x}_{k} + S^{y}_{j} S^{y}_{k} ) S^{z}_{l} \nonumber\\
&+ \sum_{I} \sum_{jkl\in\hexagon_{I}} \gamma_{r_1,r_2,r_3} ( S^{-}_{j} S^{-}_{k} S^{+}_{l} T^{+}_{I} + \mathrm{h.c.} ) \nonumber\\
&- h_{A} \sum_{I} T^{z}_{I}- h_{B} \sum_{i} S^{z}_{i},
\label{eq:heff_2mag_spin}
\end{align}
where the couplings are given by
\begin{equation}
\begin{split}
J^{xy}_{AB} &= 2 (t_{AB} + T^{AB;B}_{1} + T^{AB;B}_{2}) + T^{AB;B}_{3}, \\
J^{z}_{AB} &= U_{h,cs}, \\
J^{xy}_{BB,1} &= 2 (t_{BB,1} + T^{BB;B}_{1,1,2} + T^{BB;B}_{1,2,3})+ T^{BB;A}_{1}, \\
J^{z}_{BB,1} &= U_{BB,1},\\
J^{xy}_{BB,2} &= 2 t_{BB,2} + 2 T^{BB;B}_{2,1,3} + T^{BB;B}_{2,2,2}+ T^{BB;A}_{2}, \\
J^{z}_{BB,2} &= U_{BB,2}, \\
J^{xy}_{BB,3} &= 2 t_{BB,3} + 4 T^{BB;B}_{3,1,2} + T^{BB;A}_{3}, \\
J^{z}_{BB,3} &= U_{BB,3}, \\
J^{xy}_{AA} &= 2 t_{AA}, \\
J^{z}_{AA} &= U_{AA}, \\
h_{A} &= h + \mu_{A} + 3U_{AB} + 3U_{AA},\\
h_{B} &= h + \mu_{B} + U_{AB} + 2U_{BB,1} + 2 U_{BB,2} + U_{BB,3}.
\end{split}
\end{equation}
In the MF approximation in Sec.~\ref{sec:ss_struct},  we have neglected the small contributions coming from $T^{BB;A}_{n}$ ($n=1,2$),
for simplicity,

For the $J'=1$ case, the three strongest couplings are evaluated as
\begin{align}\label{eq:coupl_ values_spins}
(J_{AB}^{xy},J_{AB}^z)      &= ( 0.309, 0.488),\nonumber\\
(J_{BB:1}^{xy} ,J_{BB;1}^z)&=( 0.194, 0.152), \nonumber\\
(J_{BB;2}^{xy},J_{BB;2}^z) &=( 0.398, 0.638),
\end{align}
and the effective magnetic fields are
\begin{align}\label{eq:h_ values}
h_A &=h+0.337,\nonumber\\
h_B &=h+ 0.183.
\end{align}

\newpage

%%%%%%%%%%%%%%%%%%%%%%%%%%%%%%%%%%%%%%%%%%%%%%%%%%%%%%
%%%%%%%%%%%%%%%%%%%% BIBLIOGRAPHY %%%%%%%%%%%%%%%%%%%%
%%%%%%%%%%%%%%%%%%%%%%%%%%%%%%%%%%%%%%%%%%%%%%%%%%%%%%

%\bibliography{biblio}
%\bibliographystyle{./apsrev4-1}


\end{document}